\documentclass[twocolumn]{aastex631}
\usepackage{booktabs}
\usepackage{longtable}
\usepackage{amsmath}
\usepackage{xcolor}
\usepackage{array}
\usepackage{ragged2e}

\shorttitle{Evolution of the Entropy Threshold for Cooling and Feedback}
\shortauthors{Calzadilla et al.}

\begin{document}

\title{The SPT-Chandra BCG Spectroscopic Survey I: Evolution of the Entropy Threshold \\for Cooling and Feedback in Galaxy Clusters Over the Last 10 Gyr}

\correspondingauthor{Michael Calzadilla}
\email{msc92@mit.edu}

\author[0000-0002-2238-2105]{Michael S. Calzadilla}
\affil{Kavli Institute for Astrophysics and Space Research, Massachusetts Institute of Technology, Cambridge, MA 02139, USA}

\author[0000-0001-5226-8349]{Michael McDonald}
\affil{Kavli Institute for Astrophysics and Space Research, Massachusetts Institute of Technology, Cambridge, MA 02139, USA}




\author[0000-0002-5108-6823]{Bradford A. Benson}
\affil{Kavli Institute for Cosmological Physics, University of Chicago, 5640 South Ellis Avenue, Chicago, IL 60637, USA}
\affil{Fermi National Accelerator Laboratory, MS209, P.O. Box 500, Batavia, IL 60510, USA}
\affil{Department of Astronomy and Astrophysics, University of Chicago, 5640 South Ellis Avenue, Chicago, IL 60637, USA}

\author[0000-0001-7665-5079]{Lindsey E. Bleem}
\affil{High Energy Physics Division, Argonne National Laboratory, 9700 S. Cass Avenue, Argonne, IL 60439, USA}
\affil{Kavli Institute for Cosmological Physics, University of Chicago, 5640 South Ellis Avenue, Chicago, IL 60637, USA}

\author[0000-0003-2402-9003]{Judith H. Croston}
\affil{School of Physical Sciences, The Open University, Walton Hall, Milton Keynes, MK7 6AA, UK}

\author[0000-0002-2808-0853]{Megan Donahue}
\affil{Michigan State University, Physics and Astronomy Dept., East Lansing, MI 48824-2320, USA}





\author[0000-0002-3398-6916]{Alastair C. Edge}
\affil{Department of Physics, University of Durham, South Road, Durham, DH1 3LE, UK}

\author[0000-0003-4175-571X]{Benjamin Floyd}
\affil{Faculty of Physics and Astronomy, University of Missouri -- Kansas City, 5110 Rockhill Road, Kansas City, MO 64110, USA}

\author[0000-0002-7371-5416]{Gordon P. Garmire}
\affil{Huntingdon Institute for X-ray Astronomy, LLC, Huntingdon, PA 16652, USA}

\author[0000-0001-7271-7340]{Julie Hlavacek-Larrondo}
\affil{Département de Physique, Université de Montréal, Succ. Centre-Ville, Montréal, Québec, H3C 3J7, Canada}

\author[0000-0002-8314-9753]{Minh T. Huynh}
\affil{CSIRO Space and Astronomy, PO Box 1130, Bentley WA 6102, Australia}

\author[0000-0002-3475-7648]{Gourav Khullar}
\affiliation{Department of Physics and Astronomy, and PITT PACC, University of Pittsburgh, Pittsburgh, PA 15260, USA}


\author[0000-0002-0765-0511]{Ralph P. Kraft}
\affil{Center for Astrophysics | Harvard \& Smithsonian, 60 Garden Street, Cambridge, MA 02138, USA}


\author[0000-0002-2622-2627]{Brian R. McNamara}
\affil{Department of Physics and Astronomy, University of Waterloo, Waterloo, ON N2L 3G1, Canada}
\affil{Waterloo Centre for Astrophysics, University of Waterloo, Waterloo, ON N2L 3G1, Canada}
\affil{Perimeter Institute for Theoretical Physics, 31 Caroline Street North, Waterloo, ON N2L 2Y5, Canada}

\author[0000-0003-1832-4137]{Allison G. Noble}
\affil{School of Earth and Space Exploration, Arizona State University, Tempe, AZ 85287, USA}

\author[0000-0001-5725-0359]{Charles E. Romero}
\affil{Center for Astrophysics | Harvard \& Smithsonian, 60 Garden Street, Cambridge, MA 02138, USA}

\author[0000-0002-0955-8954]{Florian Ruppin}
\affil{Univ. Lyon, Univ. Claude Bernard Lyon 1, CNRS/IN2P3, IP2I Lyon, F-69622, Villeurbanne, France}


\author[0000-0003-3521-3631]{Taweewat Somboonpanyakul}
\affiliation{Kavli Institute for Particle Astrophysics and Cosmology, Stanford University, 452 Lomita Mall, Stanford, CA 94305, USA}



\author[0000-0002-3514-0383]{G. Mark Voit}
\affil{Michigan State University, Physics and Astronomy Dept., East Lansing, MI 48824-2320, USA}


\begin{abstract}

We present a multi-wavelength study of the brightest cluster galaxies (BCGs) in a sample of the 95 most massive galaxy clusters selected from South Pole Telescope (SPT) Sunyaev-Zeldovich (SZ) survey. 
Our sample spans a redshift range of $0.3 < z < 1.7$, and is complete with optical spectroscopy from various ground-based observatories, as well as ground and space-based imaging from optical, X-ray and radio wavebands. 
At $z\sim0$, previous studies have shown a strong correlation between the presence of a low-entropy cool core and the presence of star-formation and a radio-loud AGN in the central BCG.
We show for the first time that a central entropy threshold for star formation persists out to $z{\sim}1$. 
The central entropy (measured in this work at a radius of 10 kpc) below which clusters harbor star-forming BCGs is found to be as low as $K_\mathrm{10 ~ kpc} = 35 \pm 4$ keV cm$^2$ at $z < 0.15$ and as high as $K_\mathrm{10 ~ kpc} = 52 \pm 11$ keV cm$^2$ at $z \sim 1$. We find only marginal (${\sim}1\sigma$) evidence for evolution in this threshold.
In contrast, we do not find a similar high-$z$ analog for an entropy threshold for feedback, but instead measure a strong evolution in the fraction of radio-loud BCGs in high-entropy cores as a function of redshift.
This could imply that the cooling-feedback loop was not as tight in the past, or that some other fuel source like mergers are fueling the radio sources more often with increasing redshift, making the radio luminosity an increasingly unreliable proxy for radio jet power. 
We also find that our SZ-based sample is missing a small (${\sim}$4\%) population of the most luminous radio sources ($\nu L_{\nu} > 10^{42}$ erg s$^{-1}$), likely due to radio contamination suppressing the SZ signal with which these clusters are detected.

\end{abstract}


\section{Introduction} \label{sec:intro}


Galaxy clusters, the largest gravitationally bound structures in the Universe, are intricate ecosystems that allow us to investigate numerous astrophysical processes. Central to our understanding of these systems is the hot ($T \sim 10^7$ K) intracluster medium (ICM), a diffuse gas that permeates the space between the member galaxies and emits X-rays via radiative cooling. 
In the central regions of many clusters, where this reservoir of gas is relatively colder and denser, the more frequent interactions of ICM particles increases their rate of cooling and X-ray production, and eventually creates an inward flow of material called a ``cooling flow,'' where the characteristic cooling time is much shorter than the age of the universe \citep[e.g.][]{fabian94}.
Based on the standard cooling flow model, this should lead to a substantial accumulation of cool gas over time in the cluster core, subsequently fueling prodigious star formation rates in the central dominant brightest cluster galaxy (BCG) and mass accretion onto the BCG's supermassive black hole (SMBH). Despite these theoretical expectations, actual observations reveal that the amount of cooling implied by weak soft X-ray line strengths \citep{peterson06} or in the form of stars or cold molecular gas reservoirs is orders of magnitude smaller than predicted \citep[e.g.][]{1987MNRAS.224...75J, 1989AJ.....98.2018M, 1995MNRAS.276..947A, crawford99, rafferty06, 2008ApJ...681.1035O, 2015ApJ...805..177D, mcd18}, culminating in what we know as the cooling flow problem.

The leading theory for which process must be counteracting this expected cooling is feedback from active galactic nuclei (AGN; see reviews by \citealt{m+n07,m+n12,don+voit22}). AGN are actively accreting SMBHs in the centers of galaxies, which can emit enormous amounts of energy in the form of radiation (``radiative'' or ``quasar mode'' feedback) or jetted outflows (``mechanical'' or ``radio mode'' feedback). In this self-limiting feedback mechanism, the very precipitation out of the hot ICM that forms stars in the BCG eventually feeds the central SMBH which channels its accretion energy toward heating its surroundings and preventing further cooling. Radio mode feedback is especially effective at suppressing runaway cooling on large scales as it drives powerful jets, shocks, sound waves, and turbulence \citep[e.g.][]{2001ApJ...554..261C,2017ApJ...847..106L,2002MNRAS.332..271R,2016ApJ...829...90Y,2001ApJ...549..832S,2014Natur.515...85Z,2017ApJ...845...91H,2019ApJ...871....6Y,2015MNRAS.451L..60G}. Evidence for such a tightly-regulated feedback loop can be found, for instance, in multi-wavelength observations that show a strong correlation between the work $p\Delta V$ done by radio jets as they expand against the surrounding ICM and that of the cooling luminosity of the ICM due to radiative losses \citep[e.g.][]{birzan04}. Observations also seem to imply that every cool core (CC) cluster -- clusters whose central cooling times are short compared to the age of the Universe --  hosts a radio loud AGN \citep{sun09}. Finally, studies like \citet{cavagnolo08} also provide compelling evidence that once the central entropy of the ICM falls below a critical threshold, it becomes locally unstable to multiphase cooling and triggers both star formation and AGN activity \citep[see also][]{1986MNRAS.221..377N,2005ApJ...632..821P,rafferty08,2017MNRAS.464.4360M,2017ApJ...851...66H,2018ApJ...853..177P}.

While there is a wealth of evidence to support the AGN feedback mechanism \citep[see][for a review]{fabian12}, many details remain to be worked out. Of particular relevance to this study, how the balance between cooling and feedback was established and has evolved with time is still a largely unexplored area of research. 
Only recently has a window into earlier cosmic epochs been opened, with the advent of Sunyaev-Zeldovich-based (SZ) cosmological surveys discovering thousands of distant galaxy clusters in the past decade up to redshifts of $z \lesssim 2$ \citep[e.g.][]{2010ApJ...722.1180V,bleem15,2016A&A...594A..27P,2021ApJS..253....3H}, allowing for studies of their cooling and heating properties, among others. 
Past flux-limited studies have also enabled evolutionary studies, but are often limited by their biased selection of rarer, intrinsically brighter objects with increasing redshift, which the mass-limited selection of SZ surveys does not suffer from. 
Using these large SZ samples, we have so far learned that over the past 9 Gyr (i.e. $z\lesssim1$) neither the CC fraction, nor the distributions of central cooling times and entropies of the ICM have had any significant evolution \citep[see e.g.][]{mcd13_xvp,ruppin21}. The mass, size, and metallicity of cool cores have similarly experienced no evolution \citep{mcd16_bcgs,2016ApJ...826..124M,mcd17}. Examples of extreme ICM cooling at high redshifts have also provided insight into how fast some clusters and BCGs can grow, as in the case of SPT2215 \citep{calzadilla_spt2215} and SpARCS1049 \citep{2015ApJ...809..173W,2020ApJ...898L..50H}. 
On the heating side, we have also learned that AGN feedback has been operating since at least $z \sim 1$, so it must have been established at earlier times, and that the ratio of AGN heating power to cooling luminosity has also remained relatively constant at a gentle $P_\mathrm{cav}/L_\mathrm{cool} < 1$ \citep[see][]{rafferty06,nulsen09,hl12,hl15,calzadilla_spt0528,ruppin23}. 


In this study, we use one such sample of SZ-selected galaxy clusters from the South Pole Telescope (SPT) SPT-SZ survey \citep{bleem15}. 
We focus on the higher-mass subsample that has been followed up with \textit{Chandra}, defined in \citet{mcd13_xvp,mcd14}, all of which have high-angular resolution X-ray observations in hand.
The X-ray properties of this sample have been studied in detail in works like \citet{mcd13_xvp}, \citet{mantz16}, and \citet{sanders18}, and we re-analyze these data here to measure their central entropies. Unique to this study, however, is optical spectroscopy for every BCG in the sample, to look at where the cooling out of the hot X-ray emitting ICM has concentrated into forming stars. We also utilize high-resolution radio data to determine whether the cooling flows have ultimately accreted onto the central SMBH in these BCGs and are triggering mechanical AGN feedback. This one of a kind cluster survey is currently the best and only dataset capable of answering how the largest galaxies, their central supermassive black holes, and their large-scale environments have grown and co-evolved over the past $\sim 10$ Gyr. In this first paper of a series, we introduce the sample and the new data, and address whether the well-established threshold for ICM cooling instabilities that we see in local systems \citep[e.g.][]{cavagnolo08} is still present at higher redshifts and whether it has evolved with time. In other words, we seek to address whether the conditions for cooling have evolved. 
In \autoref{sec:obs} we describe our sample in more detail, explain our BCG selection process, and how we reduced all of our multi-wavelength data. 
Section \autoref{sec:entropy_cooling} connects our new star formation measurements to the central ICM entropies and discusses whether an entropy threshold exists at higher redshifts and if it has evolved with time, and compare to previous efforts to study this effect. We further discuss the connection between the entropy threshold and feedback in \autoref{sec:entropy_feedback}, as well as discuss potential sources of bias.
Finally, we summarize our takeaway results and describe future work in \autoref{sec:summary}. Throughout this paper, we assume a flat $\Lambda$CDM cosmology with $H_0 = 70$ km s$^{-1}$ Mpc$^{-1}$, $\Omega_m = 0.3$, and $\Omega_{\Lambda} = 0.7$. All measurement errors are $1\sigma$ unless noted otherwise.

\bigskip
\bigskip
\section{Sample \& Observations} \label{sec:obs}
\begin{figure}
    \centering
    \includegraphics[width=\columnwidth]{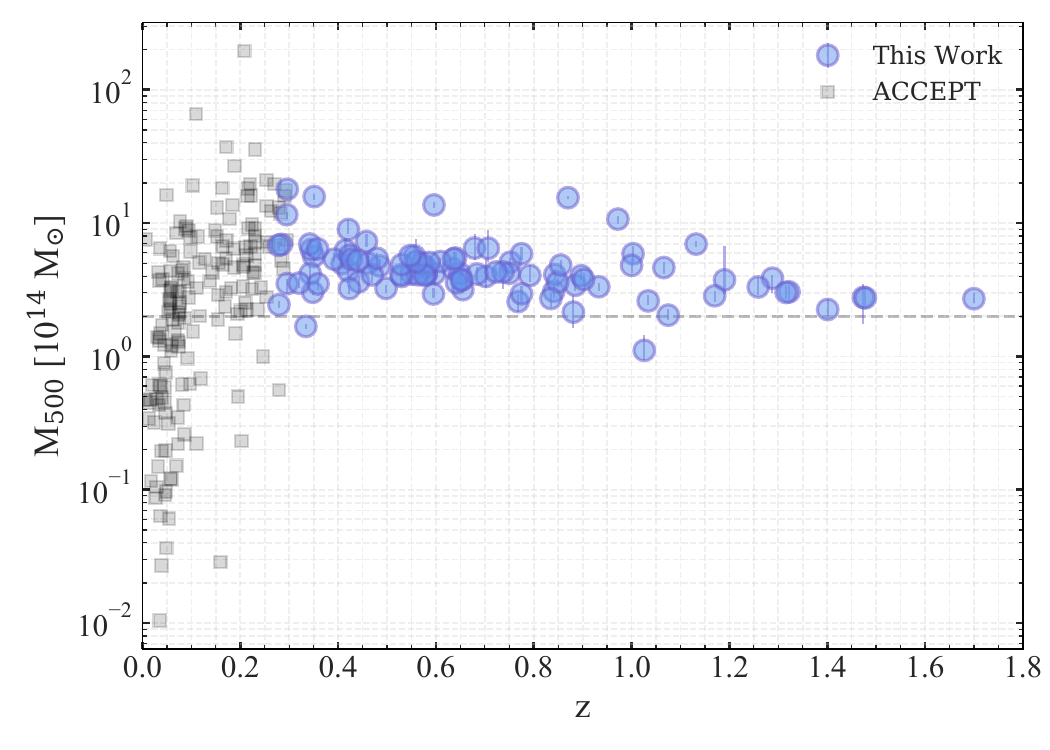}
    \caption{Mass vs redshift distribution of the data considered in this study. In blue are the clusters from our sample, which were drawn from the SPT-SZ survey, with a median mass and redshift of $\left<M_{500}\right> = 4.2 \times 10^{14}$ M$_{\odot}$ and $\left< z \right> = 0.6$. The lower mass limit of our sample  (${\sim}2 \times 10^{14}$ M$_{\odot}$) is shown as a horizontal dashed line. To look for evolutionary trends, we also compare our sample to the nearby ($0 < z < 0.3$) ACCEPT sample of galaxy clusters \citep{cavagnolo09}, shown here as grey points. For these comparisons, we enforce a minimum mass of $2 \times 10^{14}$ M$_{\odot}$ on the ACCEPT sample.}
    \label{fig:sample_M500_vs_z} 
\end{figure}

Our sample was chosen based on the ${\sim}100$ SPT-SZ selected clusters that have been followed up with \textit{Chandra}, ATCA, and Magellan observing campaigns, described further in \citet{bleem15} and in the following sections. After performing spectroscopic followup of the BCGs (with selection described in \autoref{subsubsection:bcg_selection}), some of these resulted in mis-identifications, resulting in a total remaining sample of 95 clusters. Additionally, we compare our results to the low-$z$ ACCEPT cluster sample \citep{cavagnolo09}, which we will describe in more detail in \autoref{sec:entropy_feedback}. The mass-redshift distribution for both of these samples can be seen in \autoref{fig:sample_M500_vs_z}.


\begin{figure*}
    \centering
    \includegraphics[width=\textwidth]{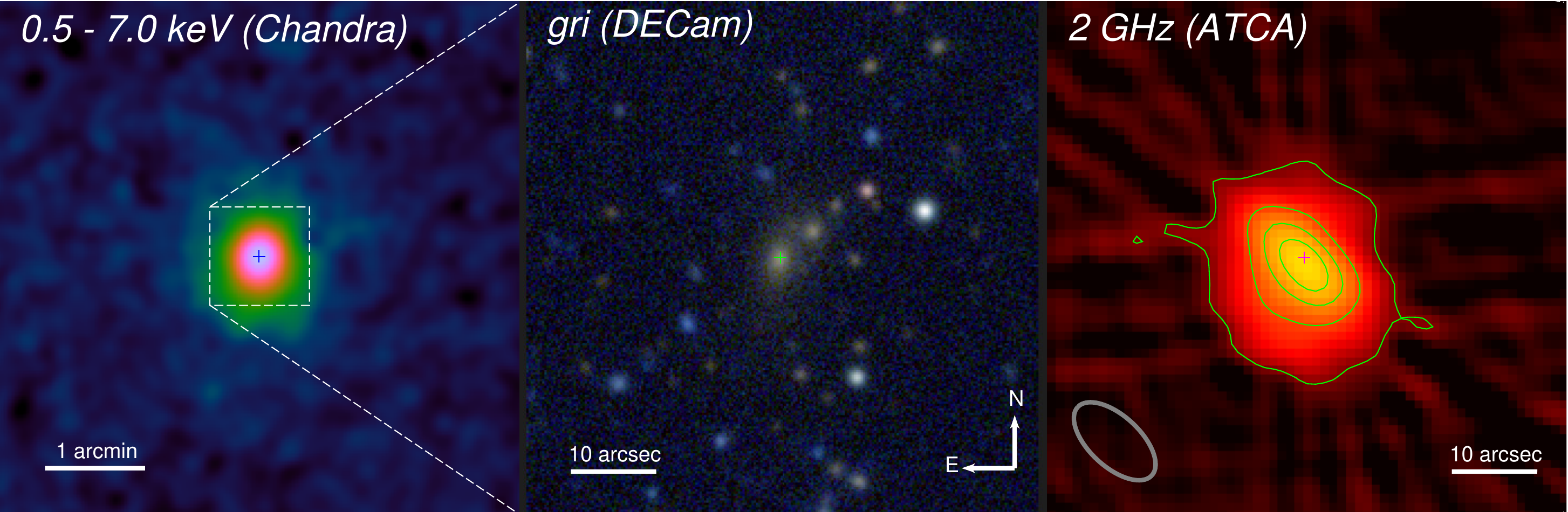}\caption{Multi-wavelength images for one of the clusters in our sample (SPT-CLJ0000-5748), showing smoothed X-ray data from \textit{Chandra} (left panel), optical data from DES (middle panel), and radio data from ATCA (right panel), with beam size shown in the bottom left. From the X-ray data, we can quantify how much hot ($T \sim 10^7$ K) gas is cooling out of the ICM to subsequently fuel both star formation onto the BCG that ionizes warm ($T \sim 10^4$ K) gas at optical wavelengths, as well as accretion onto the BCG's central AGN which we can see at radio wavelengths. The BCG (selected as described in \autoref{subsubsection:bcg_selection}) is marked in all panels with a cross ($+$). ATCA contours are shown at $[6, 50, 90, 130, 170]\times \sigma_\mathrm{rms}$, where $\sigma_\mathrm{rms} \sim 32 \, \mu$Jy beam$^{-1}$.}
    \label{fig:images}
\end{figure*}

\subsection{X-ray (Chandra)} \label{subsec:xray}
As the crucial fuel source for eventual star formation and SMBH accretion, we must first examine the X-ray emitting ICM in each of the galaxy clusters in our sample. All of our systems have been observed with the \textit{Chandra X-ray Observatory}, with the bulk of the observations coming from the multi-cycle Chandra X-ray Visionary Project (XVP; PI: B. Benson) which observed the 80 most massive clusters from SPT-SZ sample above $z>0.3$, or a Cycle 16 Large Program that observed 10 SPT-selected clusters at $z>1.2$ (PI: McDonald). X-ray data for the remaining clusters in our sample were obtained either through archival data, or various smaller Guest Observer (GO; PIs: McDonald, Hlavacek-Larrondo, Mohr) or Guaranteed Time Observer (GTO; PIs: Garmire, Murray) programs. More details on these observations can be found in \citet{mcd13_xvp,mcd14}. An example X-ray image of one of the clusters in our sample can be found in \autoref{fig:images} (left panel).

To reduce and analyze these X-ray data, we used the \textit{Chandra} Interactive Analysis of Observations {\tt (CIAO) v4.14.0} software with {\tt CALDB v4.9.8}, in a standard fashion similar to \citealt{mcd13_xvp,calzadilla_spt0528,calzadilla_spt2215,ruppin21,ruppin23}. All observations were made with the ACIS-I instrument. We applied the latest gain and charge-transfer inefficiency corrections using the \texttt{chandra\_repro} script, as well as improved background screening for observations taken in the \textsc{VFAINT} telemetry mode. Flare removal from the lightcurves is done using the \texttt{lc\_clean} script, then point sources are identified with a wavelet filter decomposition using \texttt{wavdetect} \citep[e.g.][]{wavelets} and subsequently masked to produce a clean event file. Surface brightness profiles are extracted from 20 concentric circular annuli following the uniform binning scheme of \citet{mcd17} and \citet{ruppin21}, with the outer radius of the $i^\mathrm{th}$ annulus defined as:
\begin{equation}
    r_\mathrm{out,i} = (a + bi + ci^2 + di^3) \times R_{500}
\end{equation}
where $(a, b, c, d) = (13.779, -8.8148, 7.2829, -0.15633) \times 10^{-3}$, and $R_{500}$ is the radius where the mean density is 500 times the critical density of the Universe at that redshift and is obtained from \citet{bleem15}. The annuli are centered on the X-ray peak positions -- which are found according to the procedure described in \citet{ruppin21} -- so as to trace where the ICM is cooling most rapidly. The surface brightness profile is then extracted at these annuli using the \texttt{dmextract} tool over the $0.7 - 2.0$ keV energy range.

\subsubsection{Spectral Fitting}
To measure a spectroscopic temperature profile for each object, we extract spectra using \texttt{specextract} along a much coarser binning scheme with outer radii defined by $(0.1, 0.3, 0.8) \times R_{500}$ and centered on the same X-ray peak as before. Our data quality varies considerably across the sample, and this coarse binning scheme is meant to optimally sample the worst cases where observations resulted in a few hundred counts per cluster. Many of our more deeply-observed clusters have sufficient quality to do a finer radial sampling, but we opted to have a uniform binning scheme in order to avoid any resolution bias in comparing spectroscopic profiles and central quantities later on in our analysis. Each of the spectra in these three bins were modeled over the $0.7-7.0$ keV range as an optically-thin, X-ray emitting plasma in \texttt{XSPEC v3.0.9} with the \texttt{APEC/AtomDB} thermal spectral model \citep{2001ApJ...556L..91S}, in addition to \texttt{PHABS} for photoelectric absorption \citep{1983ApJ...270..119M}. We adopt \citet{1989GeCoA..53..197A} abundances for consistency with previous literature, and Hydrogen column densities from the Leiden-Argentine-Bonn survey \citep{2005A&A...440..775K}. Redshifts are fixed to the updated values from \citet{2019ApJ...878...55B}. The metallicity is fixed to a value of $0.3 Z_\odot$. 
The background is assumed to be a combination of instrumental and astrophysical backgrounds. The instrumental background is obtained by normalizing the unscaled stowed background to the count rate of the observations in the $9-12$ keV band and subtracting from the source spectra. The remaining astrophysical background is measured from an off-source region across the ACIS-I chips, and modeled as a second \texttt{APEC} component (fixed at $kT = 0.18$ keV, $Z = Z_{\odot}, z=0$) to model soft Galactic X-ray emission, as well as a \texttt{BREMSS} model (fixed at $kT = 40$ keV) to model unresolved point sources \citep[e.g.][]{2019ApJ...885...63M}, with normalizations scaled by the ratio of areas of the background to source extraction regions. 
The resulting temperature profile was fit with the analytical model from \citet{vikhlinin06}, which was first projected along the line of sight:
\begin{equation}
    kT(x) = kT_0 \frac{\left(\frac{x}{0.045}\right)^{1.9} + \frac{T_\mathrm{min}}{T_0}}{\left(\frac{x}{0.045}\right)^{1.9} + 1} \frac{1}{\left[1 + \left(\frac{x}{0.6}\right)^2\right]^{0.45}}
\end{equation}
where $x = r/R_{500}$, and $T_0$ and $T_\mathrm{min}$ are free parameters representing the mean and core ICM temperatures respectively.

To model the ICM density, we use the surface brightness profile extracted from the fine annuli defined above, which can be expressed in terms of the emission measure ${\rm EM}(r) = \int n_e(r) n_p(r) dl$: 
\begin{equation}
    S_X(\theta) = \frac{1}{4 \pi (1+z)^4} \epsilon(T,z) \times EM(r)
\end{equation}
where $z$ is the redshift, $\epsilon(T,z)$ is the temperature and redshift-dependent ICM emissivity, and $\theta = r/D_A(z)$ is the radius scaled by angular diameter distance. At each annulus, we calculate a $S_X(\theta)$ to EM conversion factor by extracting a spectrum, and modelling it with an \texttt{APEC} model, fixing the temperature to the interpolated best-fit \citet{vikhlinin06} model at this radius, allowing only normalization to vary. The \texttt{APEC} normalization is defined as 
\begin{equation}
    N = \frac{10^{-14}}{4 \pi [D_A(z) (1+z)]^2} \int n_e n_p dl d\Omega
\end{equation}
where $D_A(z)$ is the angular diameter distance, $dl$ is the line of sight differential, $d\Omega$ is the solid angle differential. Because the \texttt{APEC} normalization contains the EM term, we can use the surface brightness profile $S_X(\theta)$ to compute an EM profile. We can arrive at the electron density ($n_e$) profile by projecting the analytic density profile from \citet{vikhlinin06} onto the plane of the sky to fit this EM profile:
\begin{equation}
    n_e(r) = n_{e0} \frac{\left(r/r_c\right)^{-\frac{\alpha}{2}}}{\left[1 + \left(r/r_c\right)^2 \right]^{\frac{3 \beta}{2} - \frac{\alpha}{4}}} \frac{1}{\left[1 + \left(r/r_s\right)^\gamma \right]^{\frac{\epsilon}{2 \gamma}}}.
\end{equation}
which is a modified beta model with a cusp, rather than a flat core (defined by $n_{e0}$, $r_{\rm c}$, $\alpha$, $\beta$), a steeper outer profile slope (defined by $r_{\rm s}$, $\gamma$, and $\epsilon$). The fitting procedure is done with the \texttt{emcee} \citep{emcee} MCMC ensemble sampler python package, as described in \citet{ruppin21}. 

\begin{figure}
    \centering
    \includegraphics[width=\columnwidth]{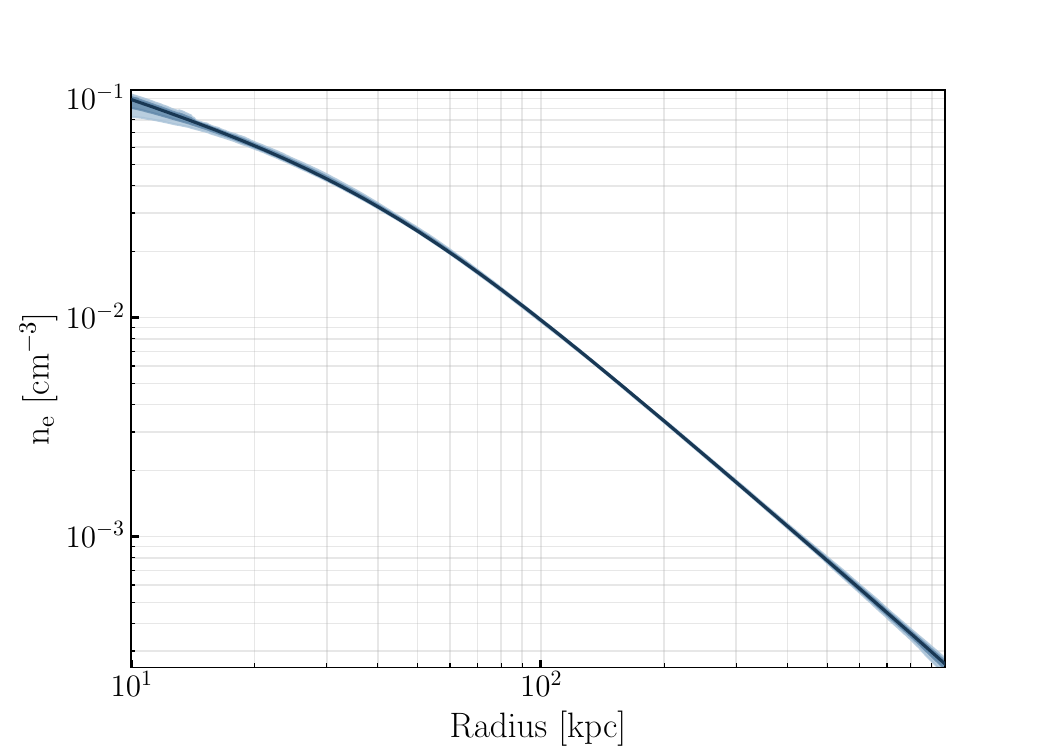}
    \includegraphics[width=\columnwidth]{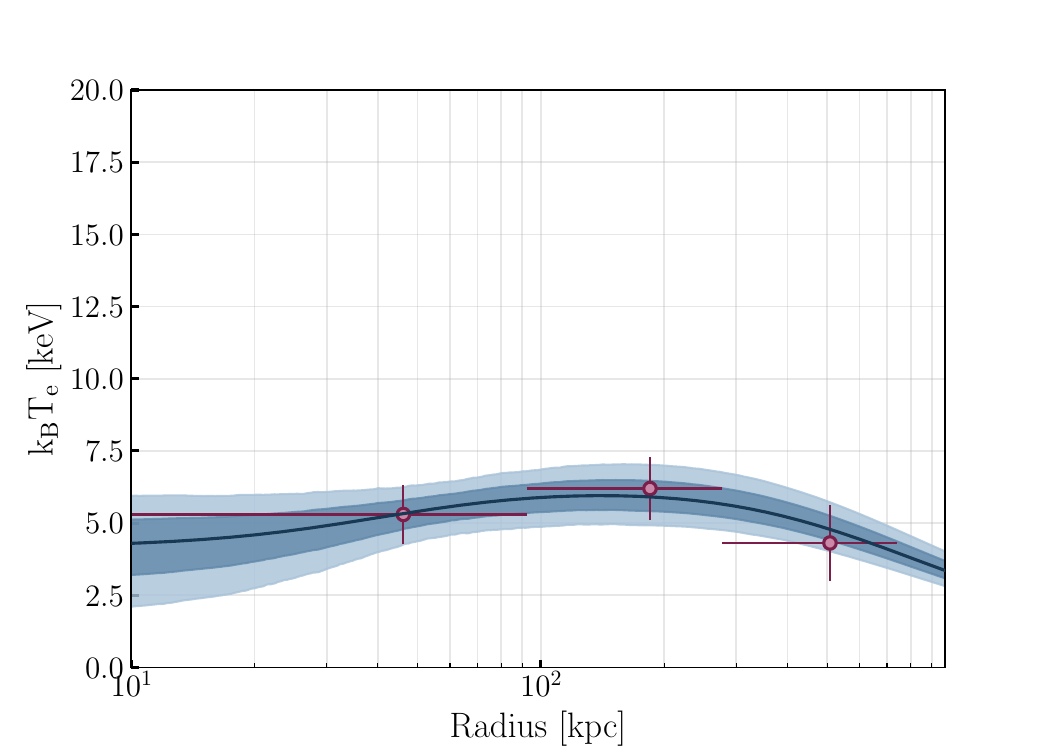}
    \includegraphics[width=\columnwidth]{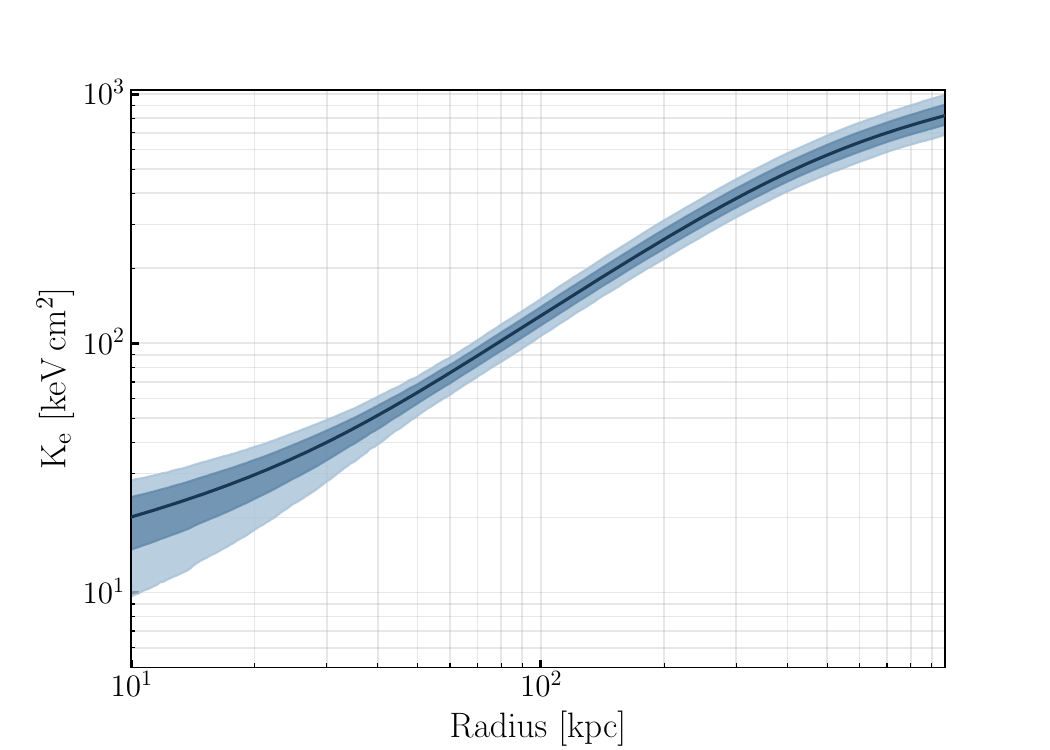}
    \caption{Thermodynamic profiles for one of the clusters in our sample (SPT-CLJ0000-5748). Top panel: electron density ($n_e$) profile. Middle panel: temperature ($kT$) profile. Bottom: entropy ($K = n_e^{-2/3} kT$) profile. 
    In all panels, the dark and light blue bands show the 1$\sigma$ and 2$\sigma$ confidence interval, respectively, while the solid line shows the best fit. The density and entropy profiles are showing the three dimensional model, while the temperature panel is showing the projected model fit to the data.}
    \label{fig:thermo_profs}
\end{figure}

Following the calculation of our temperature and density profiles (see \autoref{fig:thermo_profs}), it is straightforward to compute additional thermodynamic profiles. Of particular relevance for our analysis, we compute the pseudo-entropy profile as $K(r) = n_e^{-2/3} kT$. As a central quantity, we quote the entropy at a radius of 10 kpc for all of our cluster X-ray profiles in \autoref{tab:bcgdata}. 
We have compared the results of this analysis to the thermodynamic profiles published for the same clusters in \citet{mcd13_xvp} and \citet{sanders18}, and find good agreement overall. In particular, the agreement with \citet{sanders18} is excellent at all radii, while our profiles disagree somewhat with \citet{mcd13_xvp} due, largely, to a different choice of center -- \citet{mcd13_xvp} use the large-scale centroid as the center, while here we use the X-ray peak. We will return to this difference in \autoref{subsec:McD16}.

\subsection{Optical/IR}
\subsubsection{BCG Selection} \label{subsubsection:bcg_selection}
To trace whether the cooling that we measure from the X-ray-emitting ICM is fueling star formation in the central BCG, we first have to identify which member galaxy is indeed the BCG. 
In the local universe, this may typically be done by eye after a galaxy cluster has had sufficient time to relax and lead to a dominant BCG. However, at higher redshifts, when clusters are still disturbed and galaxies are still colliding and accreting smaller galaxies, it is often not obvious. 
Thus, we employ a more algorithmic approach, beginning with a probability-based BCG assignment. In summary, this analysis uses the optical and near-infrared spectral energy distribution (SED), in combination with the EAZY \citep{eazy} and FAST \citep{fast} codes to produce redshift ($p(z)$) and stellar mass ($p(M_*)$) distributions for each galaxy in each cluster. These distributions are used to compute the probability that each galaxy is the most massive, based on $p(M_*)$, in the three-dimensional cluster volume based on $p(z)$. For high quality data this methodology typically assigns nearly 100\% probability to a single massive galaxy, but this probability decreases as the data quality decreases, as it should. More details on this BCG probability assignment can be found in \citet{somboonpanyakul_agn} and will be provided in Noble et al. (in prep).

After assigning these BCG probabilities to member galaxies in each cluster, we then consider a hierarchy of conditions before confirming a final BCG candidate: (1) If a candidate had a BCG probability of $p\gtrsim90\%$, we considered this sufficient after additional visual inspection. (2) If there were no obviously dominant galaxies with $p\gtrsim90\%$ or there were two or more galaxies with roughly equal BCG likelihoods, then we chose the galaxy at the center of gravitational lensing arcs, assuming these arcs are part of a larger lensing ring. These arcs are visually identified and are present throughout much of our sample, and have been the subject of many papers and follow-up observations \citep[e.g.][]{bleem15}. (3) If there were no signs of lensing, we chose the brightest galaxy near the X-ray peak positions measured in \citet{mcd13_xvp}. X-ray centroids were also available, but the X-ray peak will more closely follow where most of the cooling occurs, and can move relative to the underlying gravitational potential just as the BCG should. With this heuristic, we publish these new BCG coordinates for the entire sample in \autoref{tab:bcgdata}, and proceed to collect follow-up spectroscopy for each BCG.

\subsubsection{Spectroscopy} \label{subsubsec:spectroscopy}
To determine whether the ICM is forming stars as it cools and deposits onto the central BCG, we gather spectroscopic data to look for signatures of young stars ionizing their surroundings. 
For a large portion of the objects in our sample, archival optical spectroscopy was available from the SPT collaboration's follow-up campaigns. Most of this spectroscopy (${\sim} 60\%$) was collected prior to 2017, with the GMOS spectrograph at the Gemini Southern Telescope, as well as the FORS2 spectrograph on the Very Large Telescope. Additional details on these observations and their reduction and calibration can be found in \citet{ruel14} and \citet{bayliss16} where they were originally presented. 
Lastly, we used the spectroscopic data on the two intermediate redshift systems presented in \citet{sifon13}. 

Beyond these archival observations, we present here new spectroscopy for one-third of the objects in our sample, including a few re-observations of previously observed targets due to [O\,\textsc{\lowercase{II}}] falling in a chip gap, insufficient depth, or poorly calibrated data. For these new observations, we used the IMACS spectrograph on the Magellan/Baade telescope for 13 systems in October 2019, 7 systems in October 2021, and 3 systems in November 2021. We also used the LDSS3 spectrograph to observe 2 systems in January 2020, 2 systems in October 2020, 1 system in November 2020, 4 systems in September 2021, and 1 system in November 2021. Filters and gratings for our observations were chosen to search for the presence of redshifted [O\,\textsc{\lowercase{II}}] emission from young stars ionizing their surrounding gas. Exposure times were calculated to reach a sensitivity limit of $L_\mathrm{[O\,\textsc{\lowercase{II}}]} > 1.2 \times 10^{40}$ erg s$^{-1}$ to enable a SFR measurement down to a sensitivity of at least 1 M$_\odot$ yr$^{-1}$.

To reduce these spectroscopic data, we used the standard \texttt{pyRAF/IRAF}\footnote{\url{https://iraf-community.github.io/pyraf.html}} tools. Science and calibration arc exposures were bias and flat-field corrected using the \texttt{imred.ccdred} package. We also use the \textit{response} task from the \texttt{twodspec.longslit} to fit for the shape of the lamp spectrum in the dispersion direction before flat-fielding. For multi-object spectra, the one dimensional (1D) BCG spectrum was identified and traced in the flat-fielded science frames using the \texttt{twodspec} package \textit{apall} with background subtraction. 1D spectra were extracted from the arc frames using the same traced apertures, after which a calibration solution was calculated with the \textit{identify} task. This wavelength solution was applied to the science frames with the \texttt{onedspec} \textit{refspec} and \textit{dispcor} tasks. 
The procedure for long-slit spectra was similar, except wavelength calibration was done first with \textit{identify}, \textit{reidentify}, and \textit{fitcoords}. This wavelength solution was then applied to the two-dimensional spectrum along with a rectilinear transformation with the \textit{transform} task, with background subtraction performed with the \textit{background} task. Finally, a 1D spectrum was then extracted with the \textit{apall} task.


\subsubsection{Photometry} \label{subsubsec:photometry}
Optical/IR imaging for every cluster in the SPT sample has been obtained for the purpose of optical/IR confirmation and the assessment of photometric redshifts. The acquisition and reduction of these data, which come from a wide variety of ground-based optical telescopes, are presented in full detail in \citet{bleem15}. This work also describes the full details of our aperture photometry, star-galaxy separation, and photometric calibration.


Infrared photometry was obtained from the \textit{WISE} satellite using the W1 ($3.4\mu m$) and W2 ($4.6\mu m$) bands. We cross-matched our BCG coordinates with the AllWISE source catalog \citep{allwise}, and extracted the photometric values where there was a match within 2\arcsec.


\subsubsection{BCG Star Formation Rates} \label{subsubsec:sfrs}

\begin{figure*}
    \centering
    \includegraphics[width=\textwidth]{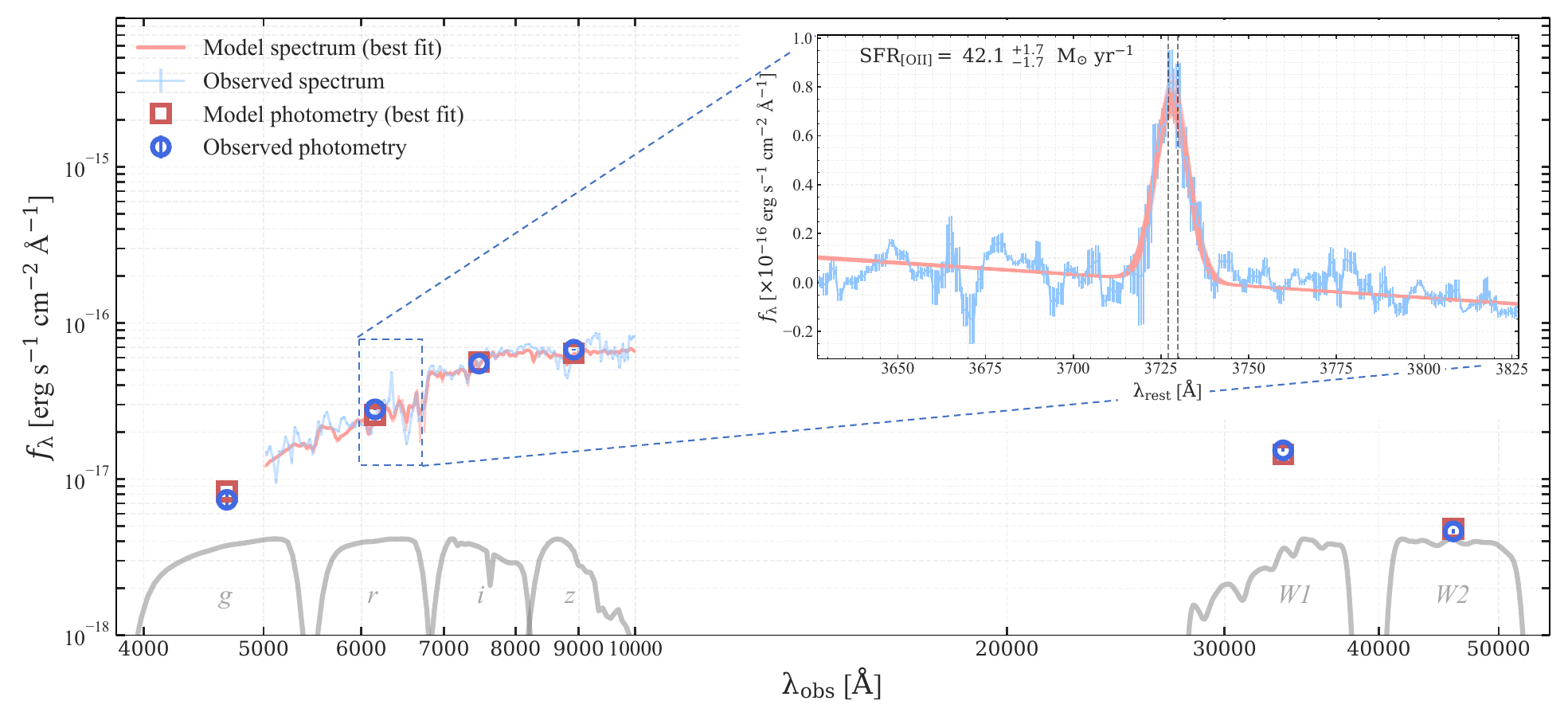}
    \caption{Spectral energy distribution (SED) fit of one of our clusters, SPT-CLJ0000-5748. Using the calibrated optical and infrared photometry of each of our clusters, we can correct for the shape of our initially uncalibrated spectra and flux calibrate using the SED fitting code \texttt{prospector}. Photometric filter bandpasses are plotted in gray and labeled. Model photometry and spectroscopy are shown in red, while observed data are shown in blue. In the inset we show our fit to the residual spectrum after subtracting the best-fit model SED from the calibrated spectrum. A linear plus Gaussian model is used to look for the presence of [O\,\textsc{\lowercase{II}}]$\lambda\lambda 3726,3729 ~ \mathrm{\AA}$ which signals star formation and is covered over our sample's entire redshift range ($0.3 < z < 1.7$).}
    \label{fig:spectral_fitting}
\end{figure*}

The wavelength-calibrated spectra obtained from the reduction steps outlined in \autoref{subsubsec:spectroscopy} were fit in combination with optical (\textit{griz}/\textit{VRI}), as well as IR WISE (\textit{W1}, \textit{W2}) photometry, using the \texttt{prospector} \citep{2021ApJS..254...22J} SED fitting code, which is especially helpful in providing spectrophotometric calibration of these uncalibrated spectra (based on calibrated photometry), constraining the underlying stellar continuum, and providing a rough estimate of the amount of intrinsic attenuation.
\texttt{Prospector} uses Markov Chain Monte Carlo (MCMC) methods to perform stellar population synthesis (SPS) based on the Python-FSPS framework \citep{2009ApJ...699..486C,2014zndo.....12157F}, incorporating the MILES stellar spectral library \citep{2011A&A...532A..95F} and MIST isochrones \citep{2016ApJ...823..102C}. Magnitudes are all converted to the AB system, and the appropriate filter throughputs corresponding to each different telescope and instrument were loaded using the python package \texttt{sedpy} \citep{sedpy}. 

In fitting our spectrophotometry, we attempt only to model the stellar continuum, and mask out telluric absorption lines as well as bright emission lines like [O\,\textsc{\lowercase{II}}], [O\,\textsc{\lowercase{III}}] $\lambda5007$, and the Hydrogen Balmer series lines H$\alpha ~\lambda 6563$ + [N II] $\lambda\lambda 6549,6585$, H$\beta ~\lambda 4861$, H$\gamma ~\lambda 4340$, and H$\delta ~\lambda 4102 $, which are typically associated with ionization from star formation and AGN \citep[e.g.][]{1998ARA&A..36..189K,kewley04}. For a given BCG, once we obtain our best-fit model, we subtract it from our flux-calibrated spectrum and look for the presence of emission lines in the residual spectrum.
To model the line-free stellar continuum, we employ a simple delayed-tau parametric star formation history of the form ${\rm SFR}(t,\tau) \propto \left(t/\tau\right) e^{-t/\tau}$, as in \citet{2022ApJ...934..177K}, with a Salpeter initial mass function \citep{1955ApJ...121..161S}, dust attenuation following \citet{calzetti00}, and additional free parameters to capture a burst of recent star formation as well. For the spectrum shape calibration, a multiplicative calibration vector describing the ratio between observed and model spectra was defined using a third order Chebyshev polynomial with \texttt{Prospector}'s \textit{PolySpecModel} class. We also 
define a noise model using a multiplicative noise inflation (i.e.\,``jitter'') term and a pixel outlier mixture model. Bayesian priors for all of the parameters described above, as well as as the free parameters of stellar mass, metallicity, age, and redshift, are reported in \autoref{tab:priors}. 

An example SED fit to spectrophotometry for one of the BCGs in our sample, SPT-CLJ0000-5748, can be found in \autoref{fig:spectral_fitting}. In this figure we show the observed photometry and the now flux- and shape-calibrated observed spectrum, along with the best-fit model to the stellar continuum and photometry from \texttt{prospector}. 
After subtracting the best-fit stellar continuum from the observed spectrum, we then proceed with fitting this residual spectrum with a linear background to model any excess continuum, plus a Gaussian to model possible [O\,\textsc{\lowercase{II}}] emission, shown in the inset in \autoref{fig:spectral_fitting}. The spectra are converted to the rest-frame using the best-fit BCG redshift.
We fit a region spanning $3727\mathrm{\AA} \pm 100\mathrm{\AA}$. A variable slope $m$ and intercept $b$, allowed to vary uniformly over the ranges $m \in [-1,1] \times10^{-16}$ erg s$^{-1}$ cm$^{-2}$ and $b \in [-5,5] \times10^{-14}$ erg s$^{-1}$ cm$^{-2}$ \AA$^{-1}$, respectively, are used to model out any residual structure to the continuum resulting from an insufficiently accurate SED fit. 
For the Gaussian component meant to capture potential [O\,\textsc{\lowercase{II}}] emission, the position is allowed to vary uniformly over the range of $3727\mathrm{\AA} \pm 31\mathrm{\AA}$ (i.e. $\pm 2500$ km s$^{-1}$), the amplitude between ${\pm}1 \times 10^{-15}$ erg s$^{-1}$ , and a velocity dispersion between $0.6 < \sigma < 6.0$ (i.e. $50-500$ km s$^{-1}$). 
These fits are performed using the ensemble MCMC sampler python package \texttt{emcee} \citep{emcee} with a Gaussian likelihood, 128 walkers, 1000 burn-in steps, and 10000 production steps. To allow for convergence, we let the chains run until they were at least 50 times the autocorrelation length, guaranteeing a sufficient number of functionally independent samples. 
An example of one of these fits performed to the residual spectrum of SPT-CLJ0000-5748 is shown in the inset of \autoref{fig:spectral_fitting}. At each iteration in the MCMC fits, a flux was calculated from integrating that specific model Gaussian. In the case of non-detections, the flux upper limit we quote is the 84\textsuperscript{th} percentile of the MCMC model fluxes.
A fit was considered a non-detection of [O\,\textsc{\lowercase{II}}] emission if the 0.15 percentiles (i.e. $-3\sigma$) of all these integrated fluxes was negative, or if the median peak model flux density was not at least $2\sigma$ above the continuum flux density adjacent to the line center. With these stringent criteria, we find that 22/95 spectra have [O\,\textsc{\lowercase{II}}] detections. Some potential detections are missed, but we opted for purity over completeness for the purposes of our analysis in the sections that follow. 

The line fluxes we measure can all be found in \autoref{tab:bcgdata}. The fluxes are converted to luminosities and then extinction-corrected using $E(B-V) = 0.32 \pm 0.13$ based on the distribution of BCG reddening measurements from \citet{crawford99}, and using the extinction law of \citet{calzetti00}, assuming $R_V = 3.1$. 
Luminosities are converted to SFRs using the SFR-L$_\mathrm{[O\,\textsc{\lowercase{II}}]}$ relation and scatter from \citet[][eqn. 4]{kewley04}.


\subsection{Radio} \label{subsec:radio}
To connect the cooling, X-ray-emitting ICM to synchrotron radiation resulting from black hole feeding in the BCG, we also gathered radio data for each of our systems. A targeted followup campaign of XVP clusters was made with the Australia Telescope Compact Array (ATCA). These clusters were initially observed at 2 GHz in January 2015, reaching an rms noise of 28-55 $\mathrm{\mu Jy}$ beam$^{-1}$. Additional, higher-resolution observations at 5 and 9 GHz were made in August 2016 to followup those systems where a strong detection was found at 2 GHz. At these frequencies, the radio maps reached an rms noise of 30-67 $\mathrm{\mu Jy}$ beam$^{-1}$ (5 GHz), and 19-77 $\mathrm{\mu Jy}$ beam$^{-1}$ (9 GHz). All ATCA observations were reduced with the 21 February 2015 release of the \texttt{Miriad} software package \citep{miriad}. 60 out of the 95 clusters in our sample had available ATCA data at any observing frequency. For the remainder of our sample, we used 887.5 MHz radio data from the Rapid ASKAP Continuum Survey\footnote{\url{https://data.csiro.au/collection/csiro:52217}} (RACS) first data release \citep{2021PASA...38...58H,2020PASA...37...48M}. RACS has worse angular resolution ($\theta \sim 25\arcsec$) than ATCA ($\theta \sim 3-6\arcsec$), but the RACS maps in \citet{2021PASA...38...58H} have uniform sensitivity (${\sim}0.3$ mJy rms) and coverage of the entire SPT-SZ footprint with a declination range of $\delta \in [-80^{\circ}, +30^{\circ}]$.

To consider a radio source a detection associated with our BCG, we required that the source position was within 5\arcsec of the BCG coordinates. For the ATCA data, the angular resolution was good enough to visually confirm these associations (see right panel in \autoref{fig:images}). These ATCA data provided 18/95 confirmed radio detections. The ASKAP data matching contained 27/95 detections, confirming 15 of the 18 ATCA detections, and found an additional 12 unique detections only for the systems that had no ATCA data (i.e. the ASKAP did not misidentify any detections where there was also higher-resolution ATCA data). Thus we find a total of 30/95 (32\%) radio detections.
The ATCA data maps were modeled interactively with a variable combination of 2D Gaussians, with major and minor axes and position angles fixed to the point spread function (psf) model parameters in each observing frequency. The sky position and amplitudes of the Gaussians were free to vary, and were estimated using an MCMC analysis. More details on this fitting procedure can be found in \citet{ruppin23}. Radio fluxes and uncertainties were estimated by sampling the posterior distributions of the model parameters at each frequency. For non-detections, we instead quoted an upper limit based on 3$\times$ the rms of the radio maps. Similarly for the ASKAP data, we quoted the integrated flux of the RACS DR1 ``continuum\_component'' source closest to our BCG within 5\arcsec, and for the non-detections we again quoted 3$\times$ the rms of the radio maps. Fluxes were converted to k-corrected 1.4 GHz rest-frame radio power assuming $L \propto \nu^{\alpha}$, and using 
\begin{equation}
   L_{1.4 ~ \mathrm{GHz},{\rm rest}} = \frac{4 \pi D_L^2(z)}{(1+z)^{1+\alpha}} \left(\frac{1.4 ~ \mathrm{GHz}}{\nu} \right)^{\alpha} F_{\nu,{\rm obs}}
\end{equation}
where $z$ is the redshift, $D_L(z)$ is the luminosity distance, and $\alpha$ is the power-law spectral index for the radio source. For ATCA data where multiple frequency bands were available (2, 5, and/or 9 GHz), the $\alpha$ was measured directly, while for the rest of the ATCA and all of the ASKAP data, we set $\alpha = -1.12\pm0.06$ from the analysis of SPT radio sources from \citet{ruppin23}.

\bigskip
\section{Redshift Dependence of the Entropy Threshold for Multiphase Cooling} \label{sec:entropy_cooling}

In the decade after the launch of \textit{Chandra}, the new era of high-resolution X-ray imaging allowed us to investigate and trace in extraordinary detail the conditions that lead to rapid cooling of the ICM. The seminal work done by \citet{cavagnolo08,cavagnolo09}, for instance, demonstrated that below a threshold of $K_0 < 30$ keV cm$^2$ in central ICM entropy, rapid cooling is simultaneously associated with high levels of H$\alpha$ emission \citep[see also][]{1986MNRAS.221..377N,2005ApJ...632..821P,rafferty08,2017MNRAS.464.4360M,2017ApJ...851...66H,2018ApJ...853..177P}, which probes multiphase cooling and is a good indicator of star formation and AGN activity \citep[e.g.][]{1998ARA&A..36..189K,kewley04}. 
This threshold is also seen in cooling time, where multiphase cooling and feedback ensues below a threshold of $t_{\rm cool} < 1$ Gyr.
\citet{mcd10} also showed that ICM gas coincident with star forming filaments observed in H$\alpha$ has cooling times that are shorter than their surroundings by roughly a factor of 4. This is strong evidence that we can link ICM cooling to the fueling of star formation, and that this same process initiates a self-regulating feedback loop in the form of feedback triggered by AGN feeding. However, given the flux-limited nature of many previous cluster surveys, these links could only be confidently claimed for nearby, low-$z$ systems for the most part. In this work, we make the first attempt to extend this line of investigation to determine whether the entropy threshold has evolved and had the same influence in the past, or in other words, whether the conditions for ICM cooling \& AGN feedback, have evolved with time.

\begin{figure}
    \centering
    \includegraphics[width=\columnwidth]{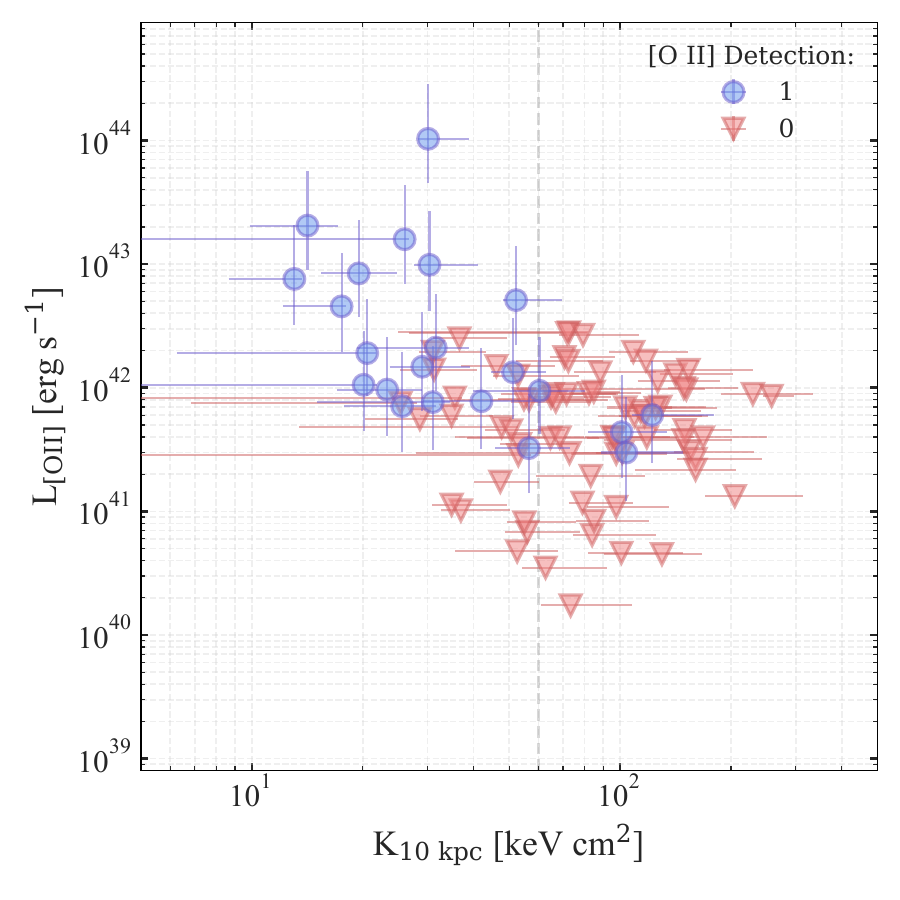}
    \caption{Central entropy of the ICM measured at a radius of 10 kpc ($K_\mathrm{10 ~ kpc}$) vs luminosity of the [O\,\textsc{\lowercase{II}}] lines ($L_\mathrm{[O\,\textsc{\lowercase{II}}]}$) measured as in \autoref{fig:spectral_fitting}. Detections are shown in blue, and upper limits are shown in red. For the first time, we show that the entropy threshold (shown here as a vertical dashed line) below which the hot ICM can condense into multiphase gas and trigger star formation persists to high ($z{\sim}1$) redshifts.}
    \label{fig:K10_vs_LOII_only}
\end{figure}


\begin{figure*}
    \centering
    \includegraphics[width=\textwidth]{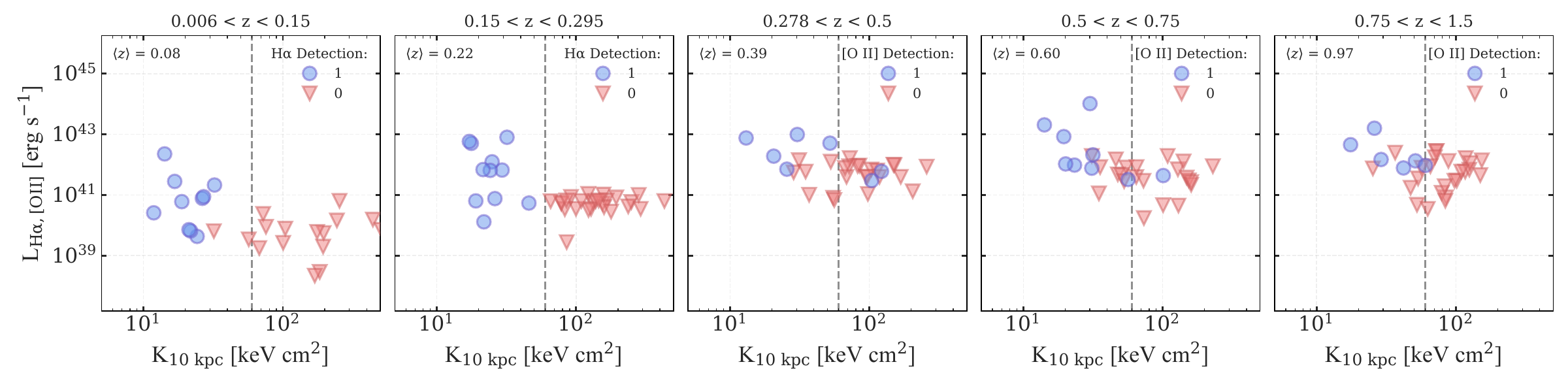}
    \caption{Central entropy ($K_\mathrm{10 ~ kpc}$) vs H$\alpha$ or $\mathrm{[O\,\textsc{\lowercase{II}}]}$ emission line luminosity (indicators of star formation). For low-$z$ clusters from the ACCEPT sample (left two panels), we show H$\alpha$ luminosities that are more readily available at low-$z$, while for high-$z$ clusters from this work (next three panels) we show $\mathrm{[O\,\textsc{\lowercase{II}}]}$ luminosities which are available for our higher-$z$ sample.
    Uncertainties are omitted for clarity. The two samples are split into various redshift bins to look for evolutionary trends in the central entropy threshold originally observed at low-$z$ (vertical dashed line). The median redshift in each bin is noted in the top left of each panel. We see that the entropy threshold of $K_\mathrm{10 ~ kpc} \approx 60$ keV cm$^2$ (vertical dashed line) persists out to our highest redshift bin.}
    \label{fig:K10_vs_LOII_1x5}
\end{figure*}

In \autoref{fig:K10_vs_LOII_only} we show the central ICM entropies as described in \autoref{subsec:xray} and measured at a radius of 10 kpc, plotted against the BCG [O\,\textsc{\lowercase{II}}] luminosities (\autoref{subsubsec:sfrs}) for our sample, which like H$\alpha$, probes thermally unstable multiphase cooling. 
We see that the vast majority of the [O\,\textsc{\lowercase{II}}] detections (and all of the strongest detections) lie to the left of a central entropy value of $K_\mathrm{10 ~ kpc} = 60$ keV cm$^2$, indicating that an entropy threshold for cooling persists in this sample spanning $0.3 < z < 1.7$ (i.e. 10 Gyr in evolution).
This persistence suggests that multiphase condensation of the ICM are already feeding star formation in central BCGs when the universe was only a few Gyr old. 


\subsection{Has the Entropy Threshold Changed With Time?}

To directly compare our analysis results with low-$z$ systems to look for evolutionary trends, we also use the same reference sample where this central entropy threshold was first measured. To do so, we re-analyzed the Archive of \textit{Chandra} Cluster Entropy Profile Tables (ACCEPT) dataset from \citet{cavagnolo09} in a manner more consistent with the data reduction and analysis steps described in \autoref{subsec:xray} to eliminate biases and allow for a more fair comparison of these low-$z$ systems with our higher-$z$ sample. The central entropy threshold $K_0$ of \citet{cavagnolo08} is somewhat model-dependent, and cannot be reliably estimated with the quality of X-ray data available for our entire sample. Instead, we choose to measure a central entropy at a projected radius of 10 kpc, which can be measured directly and coincides with a typical cooling radius for most clusters. In general, we find that $K_0$ and $K_\mathrm{10 ~ kpc}$ agree well, but the latter is far less sensitive to observation depth. In addition to this, we filter the ACCEPT sample to only include the systems whose masses are $M_\mathrm{500} > 2 \times 10^{14}$ M$_{\odot}$, which is roughly the minimum mass of our sample (see \autoref{fig:sample_M500_vs_z}).

\begin{figure}
    \centering
    \includegraphics[width=\columnwidth]{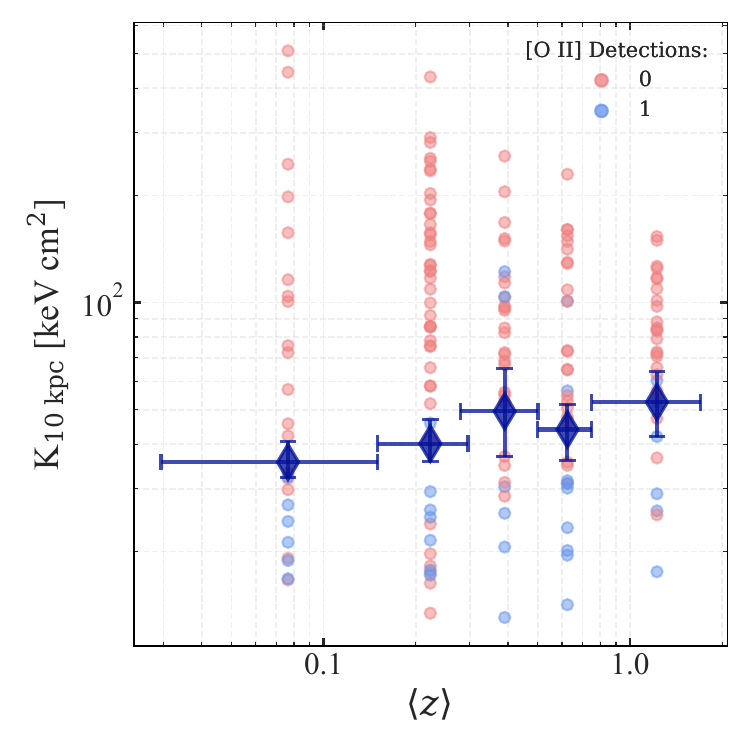}
    \caption{Bootstrapped threshold values of entropy that separate [O\,\textsc{\lowercase{II}}] detections from non-detections in each redshift bin. For each bootstrap iteration, we use a support vector classifier to find the threshold value that maximizes the margin between the two classes. We see that, while there is a slight increase in the measured threshold value with redshift, there is no strong statistical evidence for evolution.}
    \label{fig:kthresh_vs_z}
\end{figure}

While \autoref{fig:K10_vs_LOII_only} suggests that an ICM entropy threshold persists at higher redshifts, we would also like to assess if it has evolved with time. To that end, we separate our sample into three separate redshift bins with roughly equal number of clusters per bin: $0.278 < z < 0.50$, $0.50 < z < 0.75$, and $0.75 < z < 1.70$. We again plot $L_\mathrm{[O\,\textsc{\lowercase{II}}]}$ vs $K_\mathrm{10 kpc}$ for these separate redshift bins in \autoref{fig:K10_vs_LOII_1x5}. In addition to this, we plot the $L_\mathrm{H\alpha}$ data for the ACCEPT clusters, splitting into two different redshift bins of $0.003 < z < 0.15$ and $0.15 < z < 0.296$. 
We see that with increasing redshift, the $K_\mathrm{10 kpc} = 40 - 60$ keV cm$^2$ central entropy threshold identified in low-$z$ systems separates [O\,\textsc{\lowercase{II}}] detections from non-detections reasonably well as we go to higher redshifts.
To quantify this, we use a support vector classifier that maximizes the margin between two different classes ([O\,\textsc{\lowercase{II}}] detections vs non-detections in this case), with the \texttt{scikit-learn} python package \citep{scikit-learn}. Here, the margin is the distance between the threshold value and the observations closest to that threshold (i.e. the support vectors). In each redshift bin, we sample the $K_\mathrm{10 ~ kpc}$ values for each cluster within their uncertainties for 100 bootstrap iterations, each time solving for the threshold value of entropy $K_\mathrm{thresh}$ that maximizes the margin between detections and non-detections on a one-dimensional grid. Class weights are balanced proportional to the frequency of each class. 
The resulting median values and their 16\textsuperscript{th} and 84\textsuperscript{th} percentiles are sampled from the bootstrap distributions and plotted in \autoref{fig:kthresh_vs_z}. 
Plotting these threshold entropy values versus redshift for the same redshift bins as in \autoref{fig:K10_vs_LOII_1x5}, we can see that the threshold entropies increase slightly with redshift, from a minimum of $K_\mathrm{10 ~ kpc} = 35 \pm 4$ keV cm$^2$ in the lowest redshift bin ($z < 0.15$) to $K_\mathrm{10 ~ kpc} = 52 \pm 11$ keV cm$^2$ in the highest redshift bin ($0.75 < z < 1.7$). Ultimately, these values are all consistent at the ${\sim}1.5\sigma$ level, indicating that there is no strong redshift evolution in the entropy threshold for ICM cooling.

\subsection{Comparison to McDonald et al. (2016)} \label{subsec:McD16}
This work builds on the work of \citet[][hereafter M16]{mcd16_bcgs}, who looked at central entropies and SFRs for most of the BCGs in this sample. In M16, however, only 36 BCGs had optical spectroscopy, so their SFR measurements were supplemented with UV- and IR-based SFR measurements and upper limits. Such a heterogeneous dataset may introduce some systematics that are difficult to account for. 
Our sample of uniform and consistently measured [O\,\textsc{\lowercase{II}}] detections and upper limits allow for a more straightforward analysis and interpretation of some of the topics tackled in M16, and our measurements benefit from tighter measurement uncertainties and better statistics from a complete spectroscopic followup of our targets. 
One of the findings reported from M16 was that a significant number of their BCGs had SFRs $> 10$ M$_{\odot}$ yr$^{-1}$ (31/90 = 34\%), consistent with findings by \citet{2015ApJ...809..173W} and \citet{2017MNRAS.469.1259B}, and in contrast with low-$z$ systems where the occurrence rate is closer to a few percent \citep[e.g.][]{2010ApJ...715..881D,2014MNRAS.444L..63F}. 
In our study of purely [O\,\textsc{\lowercase{II}}] SFRs, we find that only 22/95 clusters have [O\,\textsc{\lowercase{II}}] detections, and only 9/95 (${\sim}9 \%$) have SFRs $\gtrsim 10$ M$_{\odot}$ yr$^{-1}$. 
This occurrence rate is in closer agreement with the $z \sim 0$ studies. Another result from M16 was that in contrast to $z \sim 0$ studies, they observed no significant correlation between BCG star formation signatures and the central ICM entropy of the host cluster. As we showed in \autoref{fig:K10_vs_LOII_only} and \autoref{fig:K10_vs_LOII_1x5}, however, we do observe a clear correlation. 

There may be many reasons for our disagreements with \citet{mcd16_bcgs} in the number of star-forming BCGs, the most likely of which being AGN contamination in the mid-IR SFRs and misidentification of BCGs based on lower quality data. The BCG selection of \citet{mcd16_bcgs} prioritized sample completeness at the expense of purity, with some BCG candidates later identified as foreground stars or galaxies.
M16 also used large-scale centroids instead of X-ray peaks for their thermodynamic profile centers, which would lead to different (often higher) central entropy measurements. 
If at higher redshifts it is more common for the X-ray peak and centroid positions to differ significantly, then these ``sloshing'' or offset cool cores
would be mislabeled as high-entropy cores in \citet{mcd16_bcgs}. 
Using the offset values from \citet{sanders18}, we find that the average angular scale difference between X-ray peak and centroid positions is about $10 \arcsec$. Given the evolution in the angular diameter distance with redshift, this angular offset corresponds to a physical offset of ${\sim}80$ kpc, which can certainly make the difference in calling a system a cool core or not. 
Alternatively, if mid-IR AGN were more common at high-$z$, which we know is the case from \citet{somboonpanyakul_agn}, then the mid-IR SFRs used in \citet{mcd16_bcgs} in the absence of optical SFRs would again wash out any sign of an entropy threshold. All of these factors likely contribute to some extent to the disagreement with our results.

\section{Redshift Dependence of the Entropy Threshold for AGN Feedback} \label{sec:entropy_feedback}

\subsection{A Disappearing Dichotomy} \label{subsec:no_radio_thresh}

\begin{figure}
    \centering
    \includegraphics[width=\columnwidth]{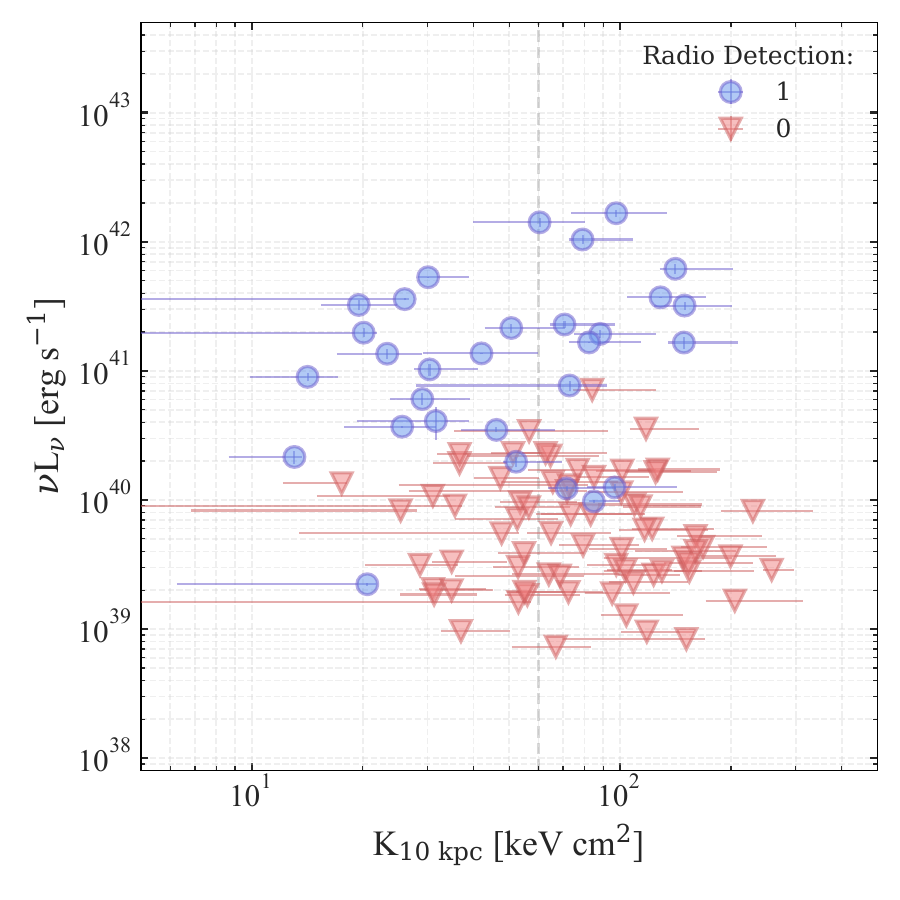}
    \caption{Central entropy of the ICM measured at a radius of 10 kpc ($K_\mathrm{10 ~ kpc}$) vs 1.4 GHz rest luminosity of the central radio source ($\nu L_{\nu}$) whenever one was associated with the BCG (i.e. a detection, in blue) or of the radio map rms noise (i.e. upper limits, shown in red). In contrast to low-$z$ systems, or like the case for star formation (\autoref{fig:K10_vs_LOII_only}), in our sample we do not see a correlation between a cluster's central entropy and AGN activity. This could be due to an actual evolution in the entropy threshold, due to a less tight feedback-cooling cycle or higher mergers at high-$z$ competing with cooling flows to drive more gas towards the AGN, for instance; or, an artificial evolution due to some biases we discuss in \autoref{subsec:radio_bias}.}
    \label{fig:K10_vs_Lradio_only}
\end{figure}

As mentioned above, the work by \citet{cavagnolo08} showed that the ICM becomes thermally unstable to cooling when its central entropy drops below some threshold $K_0$. This localized cooling not only fuels star formation, but it also triggers AGN activity by feeding the central BCG's SMBH. \citet{sun09} also showed that at low-$z$, all strong radio BCGs (defined as having $\nu L_{\nu} > 10^{39.5}$ erg s$^{-1}$) have short central cooling times (${<}4$ Gyr) making the connection between cooling and feedback quite clear. With our sample, we can investigate whether this link is still present at higher redshifts. In \autoref{fig:K10_vs_Lradio_only}, we show our central entropies measured at a radius of 10 kpc ($K_\mathrm{10 ~ kpc}$) now versus the 1.4 GHz rest-frame luminosities of the central radio sources (see \autoref{subsec:radio}). 
One of the striking features of these new data is that there is no apparent threshold between radio source detections and non-detections, in contrast to the optical luminosities associated with star formation of \autoref{fig:K10_vs_LOII_only}. On aggregate, the connection between thermally unstable cooling and AGN feedback seen at lower redshifts appear to be gone at higher redshifts. This finding is consistent with that of \citet{birzan17}, who found no evidence for a separation between cooling flow (CF) clusters and non-CF clusters based on the central radio source's luminosity at $z > 0.6$.

\begin{figure*}[ht]
    \centering
    \includegraphics[width=\textwidth]{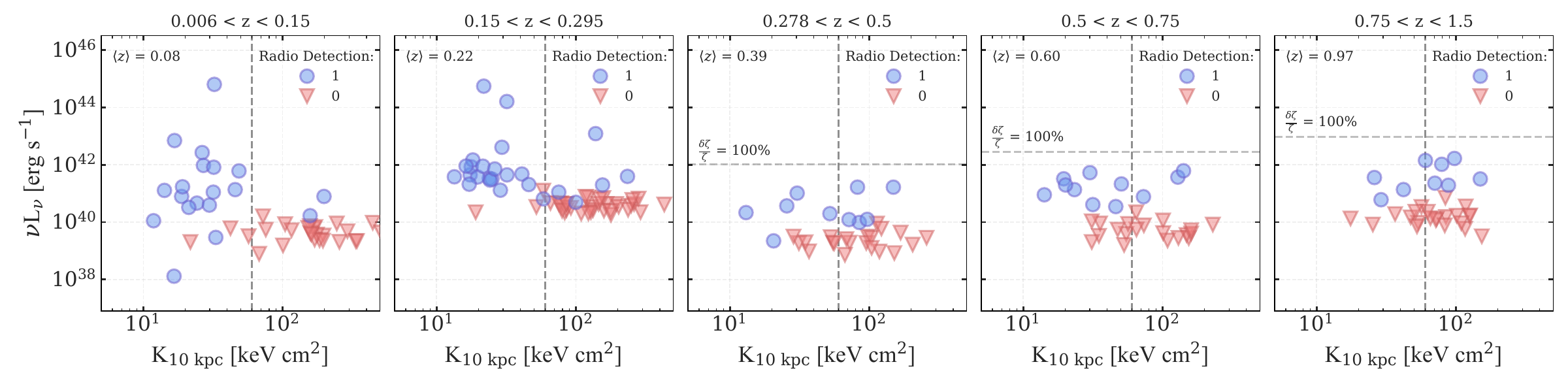}
    \caption{Same as in \autoref{fig:K10_vs_Lradio_only}, but separated into redshift bins and compared to ACCEPT clusters with $M_{500} \gtrsim 2 \times 10^{14}$ M$_{\odot}$ as before in \autoref{fig:K10_vs_LOII_1x5}. Error bars are omitted for clarity. Just as before, there is no apparent correlation between a central entropy threshold and radio AGN activity. In the low-z ACCEPT clusters, the demarcation between radio detection and non-detection is clear at $K_\mathrm{10 ~ kpc} = 60$ keV cm$^2$ in the lowest redshift bin. A higher proportion of radio detections is above the threshold in the $0.15 < z < 0.295$ redshift bin, but it is clear that the entropy distribution is still bimodal. In our SZ-based sample, the correlation vanishes abruptly. One possible source of bias is contamination to the SZ signal by the radio source, potentially making our sample deficient in high power radio sources. In the right three panels, a horizontal dashed line shows the level at which radio emission from a  $4.2 \times 10^{14}$ M$_{\odot}$ cluster at each median redshift (shown in the upper left of each panel) would suppress 100\% of the SZ signal. The varying strength of this effect of this is explored in more detail in \autoref{subsec:radio_bias} and \autoref{fig:Lradio_vs_z}.}
    \label{fig:K10_vs_Lradio_1x5}
\end{figure*}

To see whether the disappearance of the entropy threshold for feedback is a gradual one or not, we again split our sample into individual redshift bins and compare to the ACCEPT data from \citet{cavagnolo09} in \autoref{fig:K10_vs_Lradio_1x5}. The ACCEPT sample used radio luminosities from the NRAO VLA Sky Survey \citep[NVSS;][]{NVSS} and the Sydney University
Molonglo Sky Survey \citep[SUMSS;][]{SUMSS}. In the ACCEPT data, there is a clear dichotomy between radio detections and non-detections on either side of the $K_\mathrm{10 ~ kpc} = 60$ keV cm$^2$ boundary. However, this dichotomy seems to grow weaker with redshift, as roughly ${\sim}$10\%  of the detections shown in the lowest redshift bin of $0.006 < z < 0.15$ (2 out of 19) have an entropy above this threshold, while ${\sim}20$\% (5/24) of the detections are above the same threshold in the $0.15 < z < 0.295$ redshift bin. The proportion of radio detections above the threshold is approximately 50\% in the next redshift bins associated with our higher redshift SPT sample, abruptly washing out any sign of an entropy threshold.

This strong redshift evolution in the proportion of radio detections above and below the central entropy threshold seen in \autoref{fig:K10_vs_Lradio_1x5} could be interpreted in a number of ways. First, if this is a real astrophysical effect, it could imply that the cooling-feedback cycle observed in nearby systems was not as tightly-regulated in the past. 
For instance, in some cases, higher-redshift systems may often exhibit cooling that is offset from a BCG, without the opportunity for AGN accretion and feedback \citep[e.g.][]{2020ApJ...898L..50H,2019MNRAS.487.1210T}.
The more chaotic environments of high-redshift galaxy clusters that are still assembling via mergers could naturally result in more chaotic pathways of directing gas towards accreting central black holes. As the fraction of galaxies participating in mergers increases with redshift \citep[e.g.][]{2013ApJ...779..138B,2013ApJ...768....1M}, this activity is capable of driving gas towards the AGN and causing them to accrete, which would confound the specific correlation between cooling flows and AGN accretion that would otherwise result in a clean dichotomy in central entropy. 
Another related explanation could be that the lack of a dichotomy is an effect of misinterpreting the source of radio emission we measure. Recent results are making it clear that radio power becomes a progressively worse indicator of jet power at high-$z$, and an object's radio loudness is not necessarily due to the presence of jets. For instance, \citet{birzan17} find an evolution in the radio luminosity function with redshift \citep[see also][]{2017MNRAS.464.4360M,2019A&A...625A.111B}, suggesting that higher redshift radio sources are more typically associated with high-excitation radio galaxies which accrete more efficiently. In other words, at higher redshifts, a radio source is more often in quasar mode than in jet mode, which is consistent with recent studies \citep[e.g.][]{2013MNRAS.431.1638H,2013MNRAS.432..530R,somboonpanyakul_agn}, and we must be more thoughtful about the measurements we use for comparison to account for this. Studying a radio source's SED with multiple observing frequency bands would also allow for the decomposition of different contributing components like radio cores and jets. These studies will be made possible for the southern hemisphere SPT sources with upcoming ASKAP data releases, for instance \citep[see][]{2023PASA...40...34D}. 
All of these sources of confusion mentioned above are probably contributing to the lack of an observed entropy threshold for feedback to some degree.

\subsection{A Lack of Extremely Luminous Radio Galaxies in SPT Clusters} \label{subsec:radio_bias}

In addition to the lack of an observed higher-$z$ entropy threshold predicting the occurrence of radio sources as in low-$z$ systems, we also find a lack of very strong radio sources. This is qualitatively consistent with earlier findings based on ROSAT cluster surveys using a more limited redshift range \citep{2011ApJ...731...31S}.
One possible explanation for this dearth is that high mass clusters simply do not host high radio power sources as often as low-mass ones. In the ACCEPT data shown in the left two panels of \autoref{fig:K10_vs_Lradio_1x5}, we note that the highest radio powers are associated with lower-mass clusters (e.g. Hydra A, 3C388, Abell 2597), which are removed from the comparison sample when applying our mass cut of $2\times10^{14}$ M$_{\odot}$. However, even after applying a mass cut to the ACCEPT data, there remain clusters at $z<0.3$ that have central AGN with $\nu L_{\nu} > 10^{42}$ erg s$^{-1}$, including Cygnus A, Hercules A, 4C55.16, Abell 1942, PKS0745-191, Abell 2390, Abell 1361, and RXCJ0331.1-2100. These 8 clusters represent roughly 7\% of the population in this mass and redshift regime. However, it is worth noting that the ACCEPT survey is highly biased: 51\% of the clusters in the ACCEPT sample are cool cores, compared to $\sim$30\% in SZ-selected samples \citep[e.g.,][]{2017ApJ...843...76A,ruppin21}. Correcting for this, we would expect the underlying population of high-radio-luminosity BCGs to be closer to $\sim$4\% of the total population in high mass clusters.

\begin{figure}
    \centering
    \includegraphics[width=\columnwidth]{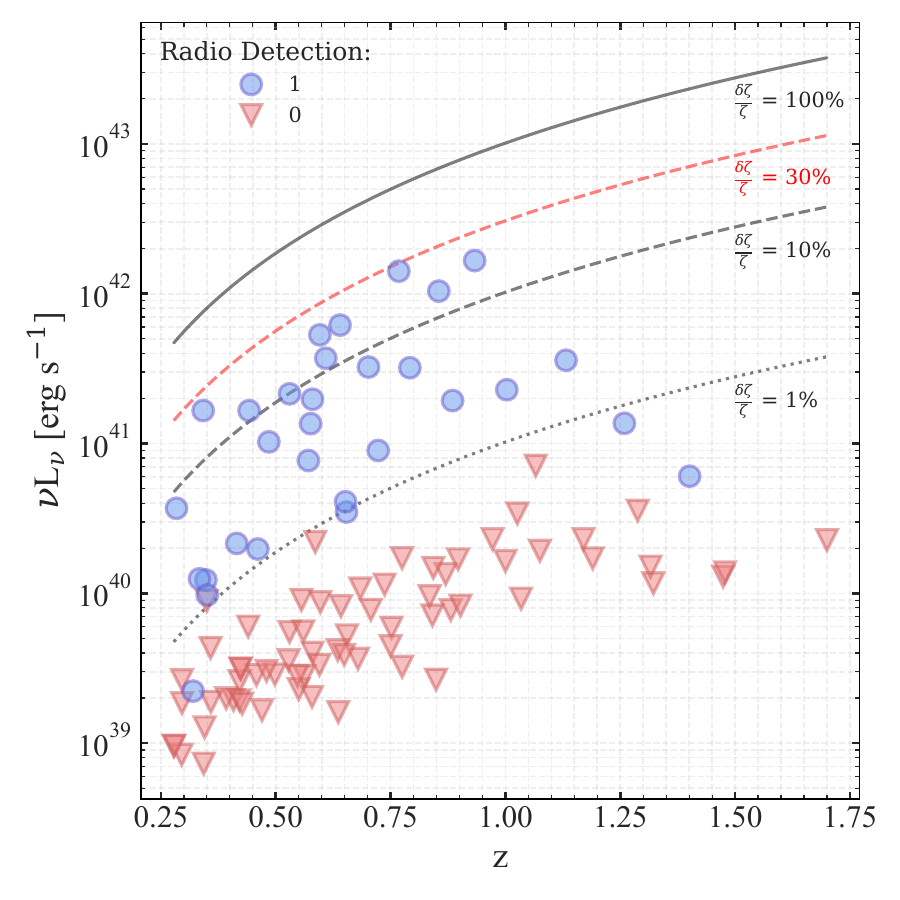}
    \caption{1.4 GHz radio luminosity vs redshift. Lines of constant flux are overlaid, depicting the amount of bias ($\delta \zeta/\zeta$) by which the radio flux suppresses the SZ signal by 1\%, 10\%, and 100\% \citep{lin15}. We see that our sample contains no clusters for which the SZ signal in a M$_{500} = 4.2 \times 10^{14}$ M$_{\odot}$ cluster would be biased by ${>}$30\%, which could be indicating a significant source of selection bias for SZ-based surveys.}
    \label{fig:Lradio_vs_z}
\end{figure}

Why are these extremely high radio luminosity systems missing in our sample? One potential concern that we consider is that the SZ surveys are biased against BCGs with extremely high radio luminosities. With the broadband spectrum from the synchrotron radiation associated with accreting SMBHs, the flux from a higher power radio source is capable of contaminating the SZ signal at the same frequencies used to detect these clusters. The SZ decrement could in effect be ``filled in'' by excess radio emission, making the SZ detections less significant. How much of an effect radio emission from a BCG has on the observed SZ signal has been quantified in various studies \citep[e.g.][]{2013ApJ...764..152S,lin15,hogan15b,2021MNRAS.508.2600D,rose22}. 
In the horizontal dashed lines plotted in  \autoref{fig:K10_vs_Lradio_1x5}, we use equation 8 from \citet{lin15} to show the level at which radio bias ($\delta \zeta/\zeta$) is 100\% of the SZ signal for a $M_{500} = 4.2 \times 10^{14}$ M$_{\odot}$ cluster (the median mass of our cluster sample) at the median redshift in each column. Above this bias threshold we do not expect to see any SZ-selected clusters. We do not plot these thresholds for the ACCEPT data as these clusters are not SZ-selected and would not suffer from the same type of selection bias. 
To investigate this further, we also plot what the radio bias at various levels ($\delta \zeta/\zeta = \{1\%, ~10\%, ~30\%, ~100\%\}$) looks like as a finer function of redshift in \autoref{fig:Lradio_vs_z}, again at a fixed median cluster mass of $4.2 \times 10^{14}$ M$_{\odot}$. 
At all redshifts, we find a lack of clusters above the line indicating a bias of 30\%.
Above a luminosity of $\nu L_{\nu} = 10^{42}$ erg s$^{-1}$, this should only result in an absence of ${\sim}$4\% of systems in our sample selection, as we estimated above from the low-$z$ ACCEPT data.  
We seem to also have a scarcity of clusters at $z>1$ and $\nu L_{\nu} \gtrsim 10^{41}$ erg s$^{-1}$, an effect that cannot be attributed to a bias in the SZ survey.

It is important to stress that this is only an attempt to quantify our lack of extremely powerful radio sources, and that this radio contamination is \textit{not} capable of contributing to our lack of an observed dichotomy between radio detections and non-detections discussed in \autoref{subsec:no_radio_thresh}. Even restricting the ACCEPT data in \autoref{fig:K10_vs_Lradio_1x5} to systems with $\nu L_{\nu} < 10^{42}$ erg s$^{-1}$, there is still a very clear contrast above and below the central entropy threshold, unlike in the higher-$z$ data. The radio contamination bias to the SZ signal is not dependent on entropy, thus we should still expect to see a dichotomy in \autoref{fig:K10_vs_Lradio_only} and \autoref{fig:K10_vs_Lradio_1x5} if one existed, assuming only that the 1.4 GHz radio luminosity is a good proxy for the AGN jet power.

Alternatively, our sample may just be missing clusters with extremely high radio luminosities because our roughly constant mass versus redshift selection (see \autoref{fig:sample_M500_vs_z}) singles out systems at higher redshifts that are progenitors of increasingly rarer systems with no low-$z$ equivalents. An evolution in the entropy threshold for feedback may look different when considering a sample of clusters along a common evolutionary track, such as the one considered in \citet{ruppin21,ruppin23}. Spectroscopic followup on such a descendent-antecedent sample would be a promising path forward.

\section{Summary \& Future Work} \label{sec:summary}
We present a multi-wavelength study of the most massive SZ-selected clusters from the SPT-SZ survey, spanning ${\sim}$10 Gyr in cosmic evolution ($0.3 < z < 1.7$). Our X-ray imaging, optical spectroscopy, and radio imaging of these clusters allows us to connect the cooling out of the hot ICM to star formation and AGN activity in the central BCG. Our results can be summarized as follows:

\begin{enumerate}
    \item For the first time, we show that a central ICM entropy threshold for multiphase condensation persists over the entire redshift range of our sample ($0.3 < z < 1.7$). This entropy threshold has only previously been demonstrated for nearby ($z \lesssim 0.3$) systems. 
    \item The threshold for ICM cooling is measured at a radius of 10\,kpc for each cluster and is found to increase slightly with redshift, from a minimum of $K_\mathrm{10 ~ kpc} = 35 \pm 4$ keV cm$^2$ in the lowest redshift bin ($z < 0.15$) to $K_\mathrm{10 ~ kpc} = 52 \pm 11$ keV cm$^2$ in the highest redshift bin ($0.75 < z < 1.7$). Ultimately, these values are roughly consistent with no statistically significant evolution.
    \item In contrast, the same entropy threshold that defines which systems should have cooling that triggers star formation and AGN feedback at low-$z$ does not predict which clusters will have radio mode feedback at higher redshifts. We find a roughly equal number of radio detections above and below the $K_\mathrm{10 ~ kpc} \sim 60$ keV cm$^2$ entropy threshold. 
    \item This lack of a radio detection dichotomy with core entropy at high-$z$ could imply that the ICM cooling-AGN feedback cycle was not as tightly-regulated in the past as it is today, or perhaps a competing source of AGN fueling like mergers, which are more prevalent at higher redshifts as well as quasars. Radio luminosities may also be an increasingly poor proxy for AGN jet power at higher redshifts, where the quasar occurrence rate is higher.
    \item We also find an absence of extremely high radio luminosity ($\nu L_{\nu} > 10^{42}$ erg s$^{-1}$) sources in our sample. This is possibly due to an SZ selection bias, as we find that our sample is missing clusters where the radio suppresses the SZ signal by as little as $\gtrsim$30\%. Importantly, this effect cannot explain the absence of the radio detection dichotomy with core entropy at high-$z$.
\end{enumerate}

This rich dataset provides many exciting possibilities for studying the evolution of the ICM cooling and AGN feedback cycle in clusters. This study is the first in a series of such possible studies. 
In upcoming papers, we plan to explore whether the effectiveness of AGN feedback has changed with time, by studying the ratio of SFR to ICM cooling rate in these clusters, as has been hinted at in \citet{calzadilla22}, \citet{mcd18}, \citet{fogarty17}, etc. 
We also plan to use this dataset to investigate whether AGN feedback was more or less bursty in the past, by studying the correlation between AGN radio power and cooling luminosity as in \citet{ruppin23}, \citet{hl12,hl15}, etc. 
Finally, we will examine in further detail the growth of BCGs using our higher quality optical spectroscopy, and see whether specific SFRs of BCGs can tell us if they are fueled preferentially via mergers or residual cooling flows, as in \citet{mcd16_bcgs}.

\bigskip
{MSC acknowledges support from the NASA Headquarters under the Future Investigators in NASA Earth and Space Science and Technology (FINESST) award 20-Astro20-0037. 
MSC and MM acknowledge additional financial support for this work provided by the National Aeronautics and Space Administration through \textit{Chandra} Award Number GO0-21124A issued by the \textit{Chandra} X-ray Center, which is operated by the Smithsonian Astrophysical Observatory for and on behalf of the National Aeronautics Space Administration under contract NAS8-03060. 
The scientific results reported in this article are based on observations made by the \textit{Chandra} X-ray Observatory, and this research has made use of software provided by the \textit{Chandra} X-ray Center (CXC) in the application package, CIAO. 

The South Pole Telescope program is supported by the National Science Foundation (NSF) through award OPP- 1852617. Partial support is also provided by the Kavli Institute of Cosmological Physics at the University of Chicago. Work at Argonne National Lab is supported by UChicago Argonne LLC, Operator of Argonne National Laboratory (Argonne). Argonne, a U.S. Department of Energy Office of Science Laboratory, is operated under contract no. DE-AC02-06CH11357. Work at Fermi National Accelerator Laboratory, a DOE-OS, HEP User Facility managed by the Fermi Research Alliance, LLC, was supported under Contract No. DE-AC02-07CH11359. This paper used data gathered with the 6.5 m \textit{Magellan} Telescopes located at Las Campanas Observatory, Chile. We thank the staff of Las Campanas for their dedicated service, which has made these observations possible. PISCO observations are supported by NSF AST-1814719.}



\facilities{
    CXO, WISE. 
    NSF/US Department of Energy 10m South Pole Telescope (SPTpol). 
    Magellan 6.5m Telescopes 
    (Clay/LDSS3C, Clay/PISCO, Baade/IMACS). 
    Gemini South Telescope (GMOS-S). 
    Blanco 4m Telescope (DECam, MOSAIC-II). 
    Swope 1m Telescope (SITe3).
    }

\software{astropy \citep{astropy}, 
    numpy \citep{numpy}, 
    scipy \citep{2020NatMe..17..261V}, 
    pandas \citep{pandas}, 
    CIAO \citep{2006SPIE.6270E..1VF}, 
    XSPEC \citep{1996ASPC..101...17A}. 
    Prospector \citep{2021ApJS..254...22J},
    python-FSPS \citep{2014zndo.....12157F}, 
    SEDpy \citep{sedpy}, 
    matplotlib \citep{2007CSE.....9...90H}, 
    seaborn \citep{2021JOSS....6.3021W}, 
    jupyter/IPython Notebooks \citep{jupyter}, 
    SAOImage DS9 \citep{ds9}, 
    pyRAF \citep{2012ascl.soft07011S}, 
    Source Extractor \citep{1996A&AS..117..393B}
    }

\vspace{15mm}


\bibliography{references}{}
\bibliographystyle{aasjournal}

\appendix

\section{Tables}

\startlongtable
\begin{deluxetable*}{lccccccccc}
\centering
\tabletypesize{\footnotesize}
\tablecaption{BCG Data \label{tab:bcgdata} }
\tablewidth{0pt}
\tablehead{
\colhead{Name} &	\colhead{$z_\mathrm{cluster}$} &	\colhead{$\alpha_\mathrm{BCG}$} &	\colhead{$\delta_\mathrm{BCG}$	} & \colhead{$z_\mathrm{BCG}$} & 	\colhead{$K_\mathrm{10 ~kpc}$}  & \colhead{$L_\mathrm{1.4 ~ GHz}$} & \colhead{$f_\mathrm{[O\, \textsc{\lowercase{II}}]}$} &	\colhead{$L_\mathrm{[O\, \textsc{\lowercase{II}}]}$} & \colhead{SFR} \\
\colhead{} & \colhead{} & \colhead{} & \colhead{} & \colhead{} & \colhead{[keV cm$^2$]} & \colhead{[$10^{40}$ erg s$^{-1}$]} & \colhead{[$10^{-16}$ erg s$^{-1}$ cm$^{-2}$]} & \colhead{[$10^{40}$ erg s$^{-1}$]} & \colhead{[M$_{\odot}$ y$^{-1}$]}
}
\decimalcolnumbers
\startdata
SPT-CLJ0000-5748 & 0.702 & 0.25015 & -57.8093 & 0.701 & $20^{+5}_{-4}$ & $32\pm3$ & $8.9\pm0.4$ & $840\pm500$ & $53^{+44}_{-25}$ \\
SPT-CLJ0013-4906 & 0.408 & 3.33048 & -49.1105 & 0.410 & $55^{+20}_{-6}$ & ${<}0.19$ & ${<}0.18$ & ${<}8.2$ & ${<}0.29$ \\
SPT-CLJ0014-4952 & 0.752 & 3.70417 & -49.8852 & 0.744 & $120^{+60}_{-20}$ & ${<}0.59$ & ${<}0.3$ & ${<}59$ & ${<}2.1$ \\
SPT-CLJ0033-6326 & 0.597 & 8.471 & -63.4449 & 0.597 & $57^{+20}_{-10}$ & ${<}0.88$ & $0.51\pm0.2$ & $32\pm20$ & $2.1^{+1.7}_{-0.98}$ \\
SPT-CLJ0037-5047 & 1.026 & 9.44771 & -50.789 & 1.029 & $57^{+40}_{-20}$ & ${<}3.4$ & ${<}0.19$ & ${<}80$ & ${<}2.9$ \\
SPT-CLJ0040-4407 & 0.350 & 10.20821 & -44.1307 & 0.350 & $110^{+50}_{-10}$ & ${<}0.89$ & ${<}2.1$ & ${<}65$ & ${<}2.4$ \\
SPT-CLJ0058-6145 & 0.885 & 14.58777 & -61.7671 & 0.830 & $88^{+40}_{-10}$ & $19\pm1$ & ${<}0.46$ & ${<}130$ & ${<}4.9$ \\
SPT-CLJ0102-4603 & 0.841 & 15.67791 & -46.071 & 0.839 & $53^{+20}_{-20}$ & ${<}0.71$ & ${<}0.018$ & ${<}4.8$ & ${<}0.17$ \\
SPT-CLJ0102-4915 & 0.870 & 15.74073 & -49.272 & 0.869 & $18^{+0.5}_{-5}$ & ${<}1.3$ & $2.9\pm0.07$ & $460\pm300$ & $29^{+23}_{-14}$ \\
SPT-CLJ0106-5943 & 0.348 & 16.61974 & -59.7202 & 0.351 & $72^{+30}_{-8}$ & $1.2\pm0.05$ & ${<}3$ & ${<}89$ & ${<}3.3$ \\
SPT-CLJ0123-4821 & 0.655 & 20.79565 & -48.3563 & 0.655 & $160^{+80}_{-20}$ & ${<}0.52$ & ${<}0.19$ & ${<}27$ & ${<}0.99$ \\
SPT-CLJ0151-5954 & 1.034 & 27.84222 & -59.9054 & 1.008 & $110^{+60}_{-20}$ & ${<}0.92$ & ${<}0.14$ & ${<}59$ & ${<}2.2$ \\
SPT-CLJ0156-5541 & 1.288 & 29.03803 & -55.703 & 1.293 & $120^{+50}_{-10}$ & ${<}3.6$ & ${<}0.22$ & ${<}170$ & ${<}5.9$ \\
SPT-CLJ0200-4852 & 0.498 & 30.14204 & -48.8713 & 0.499 & $260^{+40}_{-10}$ & ${<}0.29$ & ${<}1.2$ & ${<}86$ & ${<}3.2$ \\
SPT-CLJ0205-5829 & 1.322 & 31.44879 & -58.4821 & 1.320 & $72^{+10}_{-50}$ & ${<}1.2$ & ${<}0.35$ & ${<}280$ & ${<}9.9$ \\
SPT-CLJ0212-4657 & 0.654 & 33.12022 & -46.9488 & 0.651 & $46^{+20}_{-9}$ & $3.5\pm0.2$ & ${<}1.1$ & ${<}150$ & ${<}5.5$ \\
SPT-CLJ0217-5245 & 0.343 & 34.31195 & -52.7603 & 0.341 & $67^{+20}_{-20}$ & ${<}0.073$ & ${<}2.6$ & ${<}77$ & ${<}2.8$ \\
SPT-CLJ0232-4421 & 0.284 & 38.07726 & -44.3467 & 0.289 & $26^{+6}_{-8}$ & $3.7\pm0.07$ & $6.5\pm0.6$ & $71\pm40$ & $4.5^{+3.9}_{-2.1}$ \\
SPT-CLJ0232-5257 & 0.556 & 38.20587 & -52.9532 & 0.560 & $36^{+2}_{-40}$ & ${<}0.9$ & ${<}0.89$ & ${<}83$ & ${<}3$ \\
SPT-CLJ0234-5831 & 0.415 & 38.67609 & -58.5236 & 0.414 & $13^{+0.6}_{-4}$ & $2.2\pm0.3$ & $30\pm0.3$ & $760\pm400$ & $49^{+44}_{-23}$ \\
SPT-CLJ0235-5121 & 0.278 & 38.93863 & -51.3513 & 0.279 & $120^{+40}_{-20}$ & ${<}0.096$ & ${<}2.2$ & ${<}40$ & ${<}1.4$ \\
SPT-CLJ0236-4938 & 0.334 & 39.25696 & -49.636 & 0.336 & $97^{+50}_{-20}$ & $1.3\pm0.07$ & ${<}1.5$ & ${<}40$ & ${<}1.4$ \\
SPT-CLJ0243-5930 & 0.635 & 40.86283 & -59.5173 & 0.633 & $100^{+30}_{-20}$ & ${<}0.42$ & $0.63\pm0.2$ & $44\pm30$ & $2.8^{+2.7}_{-1.3}$ \\
SPT-CLJ0252-4824 & 0.421 & 43.20825 & -48.4162 & 0.421 & $69^{+10}_{-30}$ & ${<}0.26$ & ${<}0.81$ & ${<}40$ & ${<}1.4$ \\
SPT-CLJ0256-5617 & 0.580 & 44.12593 & -56.2974 & 0.609 & $160^{+50}_{-50}$ & ${<}0.4$ & ${<}0.22$ & ${<}22$ & ${<}0.82$ \\
SPT-CLJ0304-4401 & 0.458 & 46.07024 & -44.0254 & 0.455 & $100^{+30}_{-10}$ & ${<}0.29$ & ${<}1.2$ & ${<}69$ & ${<}2.6$ \\
SPT-CLJ0304-4921 & 0.392 & 46.06729 & -49.3573 & 0.393 & $35^{+10}_{-6}$ & ${<}0.2$ & ${<}1.5$ & ${<}59$ & ${<}2.3$ \\
SPT-CLJ0307-5042 & 0.550 & 46.96054 & -50.7012 & 0.555 & $110^{+40}_{-20}$ & ${<}0.23$ & ${<}2.2$ & ${<}200$ & ${<}7.2$ \\
SPT-CLJ0307-6225 & 0.580 & 46.81977 & -62.4465 & 0.578 & $20^{+2}_{-20}$ & $20\pm2$ & $1.8\pm0.4$ & $110\pm60$ & $6.8^{+5.8}_{-3.2}$ \\
SPT-CLJ0310-4647 & 0.707 & 47.63545 & -46.7857 & 0.707 & $73^{+30}_{-10}$ & ${<}0.78$ & ${<}0.011$ & ${<}1.7$ & ${<}0.061$ \\
SPT-CLJ0313-5334 & 1.474 & 48.48543 & -53.5708 & 1.477 & $73^{+9}_{-50}$ & ${<}1.3$ & ${<}0.27$ & ${<}280$ & ${<}9.9$ \\
SPT-CLJ0324-6236 & 0.750 & 51.05101 & -62.5988 & 0.746 & $79^{+30}_{-10}$ & ${<}0.45$ & ${<}1.4$ & ${<}270$ & ${<}9.9$ \\
SPT-CLJ0330-5228 & 0.442 & 52.73718 & -52.4703 & 0.440 & $150^{+60}_{-10}$ & $17\pm2$ & ${<}1.8$ & ${<}100$ & ${<}3.5$ \\
SPT-CLJ0334-4659 & 0.485 & 53.54573 & -46.9959 & 0.485 & $30^{+10}_{-3}$ & $10\pm1$ & $26\pm0.2$ & $990\pm600$ & $63^{+53}_{-29}$ \\
SPT-CLJ0346-5439 & 0.530 & 56.7309 & -54.6486 & 0.532 & $51^{+20}_{-8}$ & $21\pm2$ & ${<}0.55$ & ${<}45$ & ${<}1.7$ \\
SPT-CLJ0348-4515 & 0.359 & 57.07136 & -45.2498 & 0.363 & $95^{+40}_{-10}$ & ${<}0.19$ & ${<}1.3$ & ${<}40$ & ${<}1.5$ \\
SPT-CLJ0352-5647 & 0.649 & 58.23958 & -56.7977 & 0.649 & $55^{+20}_{-8}$ & ${<}0.39$ & ${<}0.59$ & ${<}81$ & ${<}2.9$ \\
SPT-CLJ0406-4805 & 0.737 & 61.73024 & -48.0825 & 0.736 & $100^{+50}_{-20}$ & ${<}1.1$ & ${<}0.026$ & ${<}4.6$ & ${<}0.17$ \\
SPT-CLJ0411-4819 & 0.424 & 62.81791 & -48.315 & 0.429 & $29^{+8}_{-8}$ & ${<}0.31$ & ${<}1.1$ & ${<}56$ & ${<}2$ \\
SPT-CLJ0417-4748 & 0.579 & 64.34616 & -47.8132 & 0.581 & $31^{+10}_{-3}$ & ${<}0.2$ & ${<}2$ & ${<}200$ & ${<}7.1$ \\
SPT-CLJ0426-5455 & 0.642 & 66.51718 & -54.9253 & 0.635 & $230^{+100}_{-40}$ & ${<}0.82$ & ${<}0.67$ & ${<}89$ & ${<}3.3$ \\
SPT-CLJ0438-5419 & 0.421 & 69.57358 & -54.3223 & 0.420 & $72^{+20}_{-20}$ & ${<}0.19$ & ${<}3.6$ & ${<}160$ & ${<}6.3$ \\
SPT-CLJ0441-4855 & 0.843 & 70.44958 & -48.9234 & 0.808 & $47^{+10}_{-7}$ & ${<}1.5$ & ${<}0.068$ & ${<}17$ & ${<}0.63$ \\
SPT-CLJ0449-4901 & 0.792 & 72.28175 & -49.0213 & 0.786 & $150^{+50}_{-30}$ & $32\pm2$ & ${<}0.2$ & ${<}45$ & ${<}1.7$ \\
SPT-CLJ0456-5116 & 0.562 & 74.11716 & -51.2764 & 0.562 & $150^{+80}_{-10}$ & ${<}0.28$ & ${<}0.33$ & ${<}30$ & ${<}1.1$ \\
SPT-CLJ0509-5342 & 0.461 & 77.33914 & -53.7035 & 0.461 & $52^{+20}_{-4}$ & $2\pm0.2$ & $15\pm0.2$ & $510\pm300$ & $33^{+27}_{-15}$ \\
SPT-CLJ0516-5430 & 0.295 & 79.15568 & -54.5005 & 0.297 & $150^{+20}_{-80}$ & ${<}0.084$ & ${<}4.8$ & ${<}97$ & ${<}3.7$ \\
SPT-CLJ0522-4818 & 0.296 & 80.56489 & -48.3048 & 0.299 & $31^{+10}_{-6}$ & ${<}0.19$ & ${<}6.7$ & ${<}140$ & ${<}5$ \\
SPT-CLJ0528-5300 & 0.768 & 82.02214 & -52.9981 & 0.766 & $60^{+20}_{-20}$ & $140\pm10$ & $0.79\pm0.04$ & $94\pm60$ & $6.1^{+4.8}_{-2.9}$ \\
SPT-CLJ0533-5005 & 0.881 & 83.40337 & -50.0958 & 0.880 & $83^{+30}_{-20}$ & ${<}0.77$ & ${<}0.069$ & ${<}19$ & ${<}0.72$ \\
SPT-CLJ0542-4100 & 0.640 & 85.70855 & -41.0001 & 0.642 & $140^{+60}_{-10}$ & $62\pm5$ & ${<}0.97$ & ${<}130$ & ${<}4.6$ \\
SPT-CLJ0546-5345 & 1.066 & 86.65741 & -53.7588 & 1.064 & $84^{+40}_{-9}$ & ${<}7.1$ & ${<}0.014$ & ${<}6.5$ & ${<}0.24$ \\
SPT-CLJ0551-5709 & 0.423 & 87.89828 & -57.1412 & 0.423 & $98^{+40}_{-20}$ & ${<}0.31$ & ${<}0.23$ & ${<}11$ & ${<}0.38$ \\
SPT-CLJ0555-6406 & 0.345 & 88.85376 & -64.1057 & 0.345 & $100^{+40}_{-20}$ & ${<}0.13$ & $1.8\pm0.4$ & $30\pm20$ & $1.9^{+1.7}_{-0.88}$ \\
SPT-CLJ0559-5249 & 0.609 & 89.93006 & -52.8242 & 0.610 & $130^{+40}_{-20}$ & $37\pm0.1$ & ${<}0.62$ & ${<}70$ & ${<}2.7$ \\
SPT-CLJ0607-4448 & 1.401 & 91.89507 & -44.8041 & 1.401 & $29^{+10}_{-5}$ & $6.1\pm0.6$ & $0.29\pm0.03$ & $150\pm90$ & $9.6^{+8}_{-4.6}$ \\
SPT-CLJ0615-5746 & 0.972 & 93.96551 & -57.7802 & 0.972 & $51^{+10}_{-7}$ & ${<}2.3$ & $0.65\pm0.2$ & $130\pm80$ & $8.6^{+7.7}_{-4.1}$ \\
SPT-CLJ0616-5227 & 0.684 & 94.14205 & -52.4525 & 0.688 & $31^{+3}_{-20}$ & ${<}1.1$ & $0.89\pm0.1$ & $77\pm50$ & $4.9^{+4.5}_{-2.2}$ \\
SPT-CLJ0640-5113 & 1.316 & 100.0725 & -51.2178 & 1.317 & $85^{+40}_{-9}$ & ${<}1.5$ & ${<}0.011$ & ${<}8.4$ & ${<}0.32$ \\
SPT-CLJ0655-5234 & 0.470 & 103.9698 & -52.568 & 0.473 & $200^{+100}_{-30}$ & ${<}0.17$ & ${<}0.22$ & ${<}13$ & ${<}0.51$ \\
SPT-CLJ2011-5725 & 0.279 & 302.8624 & -57.4197 & 0.278 & $37^{+10}_{-4}$ & ${<}0.096$ & ${<}0.56$ & ${<}10$ & ${<}0.37$ \\
SPT-CLJ2031-4037 & 0.342 & 307.9719 & -40.6252 & 0.339 & $82^{+30}_{-10}$ & $17\pm0.2$ & ${<}3.3$ & ${<}90$ & ${<}3.5$ \\
SPT-CLJ2035-5251 & 0.528 & 308.7946 & -52.8564 & 0.534 & $150^{+50}_{-60}$ & ${<}0.36$ & ${<}0.46$ & ${<}38$ & ${<}1.4$ \\
SPT-CLJ2040-4451 & 1.478 & 310.2384 & -44.8594 & 1.469 & $66^{+30}_{-20}$ & ${<}1.4$ & ${<}0.086$ & ${<}89$ & ${<}3.3$ \\
SPT-CLJ2043-5035 & 0.723 & 310.8231 & -50.5923 & 0.723 & $14^{+3}_{-4}$ & $9\pm0.7$ & $20\pm0.6$ & $2000\pm1000$ & $130^{+110}_{-61}$ \\
SPT-CLJ2106-5844 & 1.132 & 316.5192 & -58.7412 & 1.131 & $26^{+0.7}_{-20}$ & $36\pm0.8$ & $5.2\pm0.2$ & $1600\pm900$ & $100^{+87}_{-49}$ \\
SPT-CLJ2135-5726 & 0.427 & 323.9147 & -57.4376 & 0.429 & $56^{+20}_{-7}$ & ${<}0.18$ & ${<}0.14$ & ${<}6.8$ & ${<}0.24$ \\
SPT-CLJ2145-5644 & 0.480 & 326.4666 & -56.7481 & 0.481 & $53^{+20}_{-8}$ & ${<}0.3$ & ${<}1.9$ & ${<}130$ & ${<}4.6$ \\
SPT-CLJ2146-4633 & 0.933 & 326.6473 & -46.5504 & 0.928 & $98^{+40}_{-20}$ & $170\pm10$ & ${<}0.09$ & ${<}29$ & ${<}1.1$ \\
SPT-CLJ2148-6116 & 0.571 & 327.1784 & -61.2795 & 0.572 & $73^{+20}_{-40}$ & $7.7\pm0.7$ & ${<}0.31$ & ${<}30$ & ${<}1.1$ \\
SPT-CLJ2218-4519 & 0.636 & 334.7467 & -45.3144 & 0.635 & $53^{+4}_{-50}$ & ${<}0.16$ & ${<}0.23$ & ${<}29$ & ${<}1.1$ \\
SPT-CLJ2222-4834 & 0.652 & 335.7112 & -48.5764 & 0.651 & $32^{+7}_{-10}$ & $4.1\pm1$ & $2.7\pm0.2$ & $210\pm100$ & $14^{+11}_{-6.3}$ \\
SPT-CLJ2232-5959 & 0.595 & 338.1409 & -59.9981 & 0.594 & $35^{+10}_{-4}$ & ${<}0.33$ & ${<}0.1$ & ${<}11$ & ${<}0.42$ \\
SPT-CLJ2233-5339 & 0.440 & 338.315 & -53.6526 & 0.439 & $120^{+60}_{-10}$ & ${<}0.6$ & $2\pm0.5$ & $60\pm40$ & $3.9^{+3.3}_{-1.8}$ \\
SPT-CLJ2236-4555 & 1.170 & 339.2176 & -45.9305 & 1.180 & $63^{+30}_{-8}$ & ${<}2.3$ & ${<}0.006$ & ${<}3.5$ & ${<}0.13$ \\
SPT-CLJ2245-6206 & 0.586 & 341.2587 & -62.1267 & 0.560 & $65^{+20}_{-30}$ & ${<}2.2$ & ${<}0.37$ & ${<}39$ & ${<}1.4$ \\
SPT-CLJ2248-4431 & 0.351 & 342.1832 & -44.5308 & 0.347 & $85^{+20}_{-3}$ & $0.98\pm0.02$ & ${<}3$ & ${<}93$ & ${<}3.4$ \\
SPT-CLJ2258-4044 & 0.897 & 344.7011 & -40.7418 & 0.897 & $100^{+50}_{-20}$ & ${<}1.7$ & ${<}0.1$ & ${<}30$ & ${<}1.1$ \\
SPT-CLJ2259-6057 & 0.855 & 344.7541 & -60.9595 & 0.788 & $79^{+30}_{-7}$ & $100\pm8$ & ${<}0.044$ & ${<}12$ & ${<}0.43$ \\
SPT-CLJ2301-4023 & 0.835 & 345.4708 & -40.3876 & 0.857 & $54^{+20}_{-6}$ & ${<}0.96$ & ${<}0.14$ & ${<}35$ & ${<}1.3$ \\
SPT-CLJ2306-6505 & 0.530 & 346.7231 & -65.0882 & 0.529 & $48^{+5}_{-30}$ & ${<}0.55$ & ${<}0.6$ & ${<}48$ & ${<}1.8$ \\
SPT-CLJ2325-4111 & 0.358 & 351.2988 & -41.2037 & 0.362 & $170^{+80}_{-20}$ & ${<}0.43$ & ${<}1.2$ & ${<}40$ & ${<}1.4$ \\
SPT-CLJ2331-5051 & 0.576 & 352.9631 & -50.865 & 0.578 & $23^{+6}_{-6}$ & $14\pm1$ & $1.7\pm0.4$ & $97\pm50$ & $6.1^{+4.7}_{-2.8}$ \\
SPT-CLJ2332-5053 & 0.560 & 353.0249 & -50.8849 & 0.579 & $65^{+30}_{-9}$ & ${<}0.56$ & ${<}0.87$ & ${<}84$ & ${<}3$ \\
SPT-CLJ2335-4544 & 0.547 & 353.7854 & -45.7391 & 0.546 & $130^{+40}_{-40}$ & ${<}0.28$ & ${<}0.051$ & ${<}4.5$ & ${<}0.16$ \\
SPT-CLJ2337-5942 & 0.775 & 354.3651 & -59.7013 & 0.779 & $130^{+60}_{-10}$ & ${<}1.7$ & ${<}0.53$ & ${<}110$ & ${<}4$ \\
SPT-CLJ2341-5119 & 1.003 & 355.3015 & -51.3291 & 1.003 & $71^{+30}_{-6}$ & $23\pm2$ & ${<}0.46$ & ${<}180$ & ${<}6.5$ \\
SPT-CLJ2341-5724 & 1.259 & 355.3533 & -57.417 & 1.257 & $42^{+20}_{-10}$ & $14\pm0.2$ & $0.19\pm0.06$ & $78\pm40$ & $4.9^{+3.9}_{-2.2}$ \\
SPT-CLJ2342-5411 & 1.075 & 355.6913 & -54.1848 & 1.081 & $37^{+10}_{-6}$ & ${<}1.9$ & ${<}0.51$ & ${<}250$ & ${<}9.1$ \\
SPT-CLJ2344-4243 & 0.596 & 356.1829 & -42.7201 & 0.596 & $30^{+9}_{-2}$ & $53\pm0.2$ & $170\pm2$ & $10000\pm6000$ & $670^{+550}_{-320}$ \\
SPT-CLJ2345-6405 & 1.000 & 356.2376 & -64.0972 & 1.127 & $120^{+60}_{-20}$ & ${<}1.6$ & ${<}0.17$ & ${<}69$ & ${<}2.5$ \\
SPT-CLJ2352-4657 & 0.902 & 358.0678 & -46.9602 & 0.908 & $25^{+3}_{-20}$ & ${<}0.83$ & ${<}0.25$ & ${<}76$ & ${<}2.6$ \\
SPT-CLJ2355-5055 & 0.320 & 358.9478 & -50.928 & 0.318 & $21^{+1}_{-10}$ & $0.22\pm0.006$ & $13\pm2$ & $190\pm100$ & $12^{+10}_{-5.6}$ \\
SPT-CLJ2359-5009 & 0.775 & 359.9324 & -50.1722 & 0.775 & $150^{+80}_{-20}$ & ${<}0.32$ & ${<}0.66$ & ${<}140$ & ${<}4.8$ \\
\enddata
\tablecomments{Column 1: SPT cluster name. Column 2: redshift from SPT-SZ catalog. Column 3, 4: R.A. and Declination of BCG. Column 5: BCG redshift as determined from SED fitting of spectrophotometry. Column 6: ICM pseudo-entropy measured at a radius of 10 kpc. Column 7: k-corrected 1.4 GHz luminosity measured from radio data. Column 8: Raw measured [O\,\textsc{\lowercase{II}}] flux that has not been corrected by extinction. Column 9: [O\,\textsc{\lowercase{II}}] luminosity measured from spectroscopic line fitting. Column 10: [O\,\textsc{\lowercase{II}}] SFR estimate, using \citet{kewley04} }
\end{deluxetable*}
\(\)

\clearpage

\begin{deluxetable}{p{0.1\linewidth} p{0.49\linewidth} p{0.3\linewidth}}
\tabletypesize{\footnotesize}
\tablecolumns{3}
\tablewidth{0pt}
\tablecaption{ Free parameters used in \texttt{Prospector} SED fitting \label{tab:priors}}
\tablehead{
\colhead{Parameter} & \colhead{Description} & \colhead{Priors}} 
\startdata
$z_{\rm obs}$        & Observed redshift, initalized to mean $z$ from \citet{2020ApJS..247...25B} & TopHat: [$z-0.05$, $z+0.05$] \\
M$_{\rm BCG}$ (M$_{\odot}$) & Total stellar mass formed & Log$_{10}$ uniform: [$10^9, 10^{13}$] \\
$\log(Z/Z_{\odot})$  & Stellar metallicity in log solar units & Clipped normal: $\mu = 0.0$, $\sigma = 0.3$, range=[$-2.0, 1.0$] \\
$t_{\rm age}$        & Age of galaxy        & TopHat: [$0$, age of universe at $z_{\rm obs}$] \\
$\tau$               & e-folding time of star formation history in Gyr & Log$_{10}$ uniform: [$0.01, 3.0$] \\
$D$                & Optical depth for stellar light attenuation by dust for old stars using extinction curve from \citet{calzetti00}, where observed flux $I = I_0 e^{-D}$ & TopHat: [$0, 1$] \\
$f_{\rm burst}$      & Fraction of total stellar mass formed in a recent star formation burst & TopHat: [$0, 0.5$] \\
$f_{\rm age, burst}$ & Time at which burst happens, as a fraction of $t_{\rm age}$ & TopHat: [$0.1, 1$] \\
$\sigma_v$           & Velocity smoothing in km s$^{-1}$ & TopHat: [$150, 800$] \\
spec$_{\rm norm}$    & Spectrum normalization factor to match photometry & Log$_{10}$ uniform: [$1\times 10^{-6}, 10.0$] \\
spec$_{\rm jitter}$  & Multiplicative noise inflation term, which inflates the noise in all spectroscopic pixels as necessary to get a statistically acceptable fit & TopHat: [$1,10$] \\
spec$_{\rm outlier}$ & Pixel outlier mixture model, to marginalize over poorly modeled noise like residual sky lines or missing absorption lines & TopHat: [$0.0001,1$] \\
($p_1$, $p_2$, $p_3$) & Continuum calibration (Chebyshev) polynomial & TopHat: n=3: [$-0.2/(n+1), 0.2/(n+1)$] \\
 & & \\
\enddata
\end{deluxetable}





\end{document}